\documentclass[12pt]{article}
\usepackage{a4wide}
\usepackage{amssymb}
\usepackage{graphicx}
\begin{document}
{\renewcommand{\thefootnote}{\fnsymbol{footnote}}
\medskip
\begin{center}
{\LARGE  Fluctuation energies in quantum cosmology}\\
\vspace{1.5em}
Martin Bojowald\footnote{e-mail address: {\tt bojowald@gravity.psu.edu}}
\\
\vspace{0.5em}
Institute for Gravitation and the Cosmos,\\
The Pennsylvania State
University,\\
104 Davey Lab, University Park, PA 16802, USA\\
\vspace{1.5em}
\end{center}
}

\setcounter{footnote}{0}

\begin{abstract}
  Quantum fluctuations or other moments of a state contribute to energy
  expectation values and can imply interesting physical effects. In quantum
  cosmology, they turn out to be important for a discussion of density bounds
  and instabilities of initial-value problems in the presence of signature
  change in loop-quantized models. This article provides an effective
  description of these issues, accompanied by a comparison with existing
  numerical results and an extension to squeezed states. The comparison
  confirms that canonical effective methods are well-suited for computations
  of properties of physical states. As a side product, an example is found for
  a simple state in which quantum fluctuations can cancel holonomy
  modifications of loop quantum cosmology.
\end{abstract}

\section{Introduction}

Fluctuation energies are well-known from the main example of the zero-point
energy, an energy contribution that depends on quantum fluctuations and is
evaluated for the ground state. Heuristically, the ground state has a
non-vanishing energy, in contrast to the classical theory, because the
uncertainty principle does not allow both the position and the momentum
fluctuation to be zero. While expectation values of position and momentum can
minimize the classical energy function, quantum fluctuations provide an
additional zero-point energy.  One can see the relationship to quantum
fluctuations clearly by writing the energy expectation value of the harmonic
oscillator, in an arbitrary state, as
\begin{equation} \label{E}
 \langle\hat{E}\rangle= \frac{\langle\hat{p}\rangle^2}{2m}+ \frac{1}{2}
 m\omega^2 \langle\hat{x}\rangle^2+ \frac{(\Delta p)^2}{2m}+
 \frac{1}{2}m\omega^2 (\Delta x)^2\,.
\end{equation}
In a stationary state, one has
$\langle\hat{x}\rangle=0=\langle\hat{p}\rangle$, and only the fluctuation
terms remain. If one inserts the ground-state values $(\Delta
x)^2=\hbar/(2m\omega)$ and $(\Delta p)^2=\frac{1}{2}m\omega\hbar$, one obtains
the zero-point energy $\frac{1}{2}\hbar\omega$.

Although such fluctuation terms can contribute to any state of a system in
quantum mechanics, they are usually not considered significant for
semiclassical states. In this article, we discuss the analogous notion for
quantum cosmology. The situation is then rather different because the systems
considered in this field do not give rise to a natural ground state. The
more-general notion of fluctuation energies is therefore preferred compared to
zero-point energies. Quantum fluctuations (or other moments of a state) may
not seem to contribute a significant amount compared with the total matter
energy contained in the universe. However, several quantum-geometry effects
have been suggested for quantizations of space-time. Some of them, for
instance in loop quantum cosmology, imply corrections to the classical
Friedmann equation which are able to cancel the classical matter energy in
certain regimes. If a fluctuation energy is left after the cancellation, it
may have some influence on the dynamics of a universe model.

The main example for such a cancellation effect is the ``bounce'' scenario
proposed in loop quantum cosmology, based on the modified (spatially flat)
Friedmann equation
\begin{equation} \label{ModFried}
 \left(\frac{\dot{a}}{a}\right)^2=\frac{8\pi G}{3} \rho\left(1-
   \frac{\rho}{\rho_{\rm QG}}\right)
\end{equation}
with a quantum-gravity parameter $\rho_{\rm QG}$ which one can think of as
being close to the Planck density \cite{AmbigConstr,APSII}. (This equation has
been derived in a reliable way for models sourced by a free, massless scalar
\cite{BouncePert}.)  Near $\rho\sim\rho_{\rm QG}$, quantum-geometry
corrections may cancel the classical term of the matter energy density. A
solution with $\dot{a}=0$ at one time then becomes possible, leading to a
turn-around of the scale factor at high density. If there is a
fluctuation-dependent contribution to (\ref{ModFried}), viewed as an effective
equation of quantum cosmology, it may become significant near
$\rho\sim\rho_{\rm QG}$ and might change the dynamics as well as the density
realized at the bounce.

As it turned out, loop quantum cosmology is not viable as a stand-alone
universe model. The modification seen in the isotropic background equation
(\ref{ModFried}) also affects equations for inhomogeneous fields. As a
consequence, one can show that the underlying space-time structure is
drastically modified compared to the classical one
\cite{ConstraintAlgebra,JR,ScalarHol,HigherSpatial}. (It remains consistent
and anomaly-free, presenting an effective model of quantum space-time.) These
effects are visible only when inhomogeneity is included in the equations in a
consistent way, paying due attention to covariance and the gauge
transformations it is related to. If one were to fix the gauge before
considering quantum corrections or to implement other restrictions of the
gauge structure, these space-time effects would be missed. Exactly homogeneous
minisuperspace models or models in which inhomogeneous modes are added on to
the minisuperspace dynamics therefore cannot be considered reliable. (See
\cite{ReviewEff,ConsCosmo} for detailed discussions.)

The main example for unexpected quantum space-time effects in loop quantum
gravity is signature change at high density \cite{Action,PhysicsToday}:
Space-time turns into a quantum version of 4-dimensional Euclidean space
before the density $\rho_{\rm QG}$ is reached. Accordingly, equations for
inhomogeneous fields in this regime are elliptic rather than hyperbolic and do
not allow well-posed initial-value problems. There is no causal structure and
no deterministic evolution through high density, implying that an
interpretation of minisuperspace dynamics as bounce models, according to
equations such as (\ref{ModFried}), is incorrect.

Nevertheless, some properties of background solutions in loop quantum
cosmology are of interest for details of the signature-change transition at
high density. In this context, we analyze fluctuation energies contributing to
(\ref{ModFried}). We present two applications in Section~\ref{s:App}: An
estimate of the severeness of instabilities of an initial-value formulation in
the Euclidean phase, and a derivation of several relationships between moments
that help to explain some features seen in numerical evolutions of wave
functions \cite{Chimera,NumBounce}. An appendix provides more-technical
details on expectation values and moments of observables in physical states.

\section{States}

For a free, massless scalar $\phi$ with momentum $p_{\phi}$, we have the
energy density
\begin{equation} \label{rhofree}
 \rho_{\rm free}= \frac{1}{2}\frac{p_{\phi}^2}{a^6}\,.
\end{equation}
The classical Friedmann equation of spatially flat models (which is
(\ref{ModFried}) for $\rho_{\rm QG}\to\infty)$ is equivalent to the
Hamiltonian constraint $C=0$ with the ``energy''
\begin{equation}
 C=-\frac{3}{8\pi G} a\dot{a}^2+H_{\rm free}= -\frac{3}{8\pi G}
 a\dot{a}^2+\frac{1}{2} \frac{p_{\phi}^2}{a^3}\,.
\end{equation}

\subsection{Dynamical equations}

In quantum cosmology, based on Dirac's formalism to implement constraints, $C$
is quantized to an operator $\hat{C}$ which is to annihilate physical states,
$\hat{C}\psi=0$, as an analog of the classical condition $C=0$. Acting on
states $\psi(a,\phi)$, on which the momentum $p_a= -(3/4\pi G) a\dot{a}$ turns
into a derivative operator, we obtain a Wheeler--DeWitt equation \cite{DeWitt}
of the form
\begin{equation} \label{WdW}
 \frac{2\pi
 G\hbar^2}{3} \left(\frac{1}{a}\frac{\partial^2\psi}{\partial a^2}
 +\frac{2}{a^2} \frac{\partial\psi}{\partial a}+ \frac{1}{4a^3}\psi\right)
+ \hat{H}_{\rm free}\psi=0
\end{equation}
with $\hat{H}_{\rm free}=\frac{1}{2}a^{-3}\hat{p}_{\phi}^2=
-\frac{1}{2}\hbar^2a^{-3}\partial^2/\partial\phi^2$. To be specific, we have
chosen one ordering of the non-commuting operators $\hat{a}$ and $\hat{p}_a$
without implying any uniqueness of this choice. As will be shown below, it
implies convenient solvability features.

In loop quantum cosmology, the derivative $\partial/\partial a$ is replaced by
a difference operator. Not only factor-ordering but also discretization
ambiguities then apear, which could be resolved only if one were able to
derive cosmological models from the full theory of loop quantum gravity
\cite{ThomasRev,Rov,ALRev}. (But even then, the full theory itself might be
subject to ambiguities.) One question is which variable should have an
equi-spaced discretization step, $a$ or some function of it. In order to
parameterize this ambiguity to some degree, we use a basic canonical pair
\begin{equation}
 |Q|:=\frac{3a^{2(1-x)}}{8\pi G (1-x)}\quad,\quad P=-a^{2x}\dot{a}\,.
\end{equation}
(The variable $Q$ can take both signs, according to the orientation of a
spatial triad. Homogeneous variables are defined by spatial integrations and
depend on the coordinate volume of the integration region. For simplicity, we
may assume this value to equal one. See \cite{Springer} for details.)  The
parameter $x$ is a label for different discretization schemes, so that a
quantization of the power $a^{2(1-x)}$ becomes equidistant. This parameter can
heuristically be related to the dynamical refinement of spatial lattices
underlying states in loop quantum gravity \cite{InhomLattice,CosConst}. The
most common choice is $x=-1/2$ for equidistant volume $V=4\pi GQ|_{x=-1/2}$
\cite{APSII}, but it is not unique. The variables $(a,p_a)$ traditionally used
in Wheeler--DeWitt quantum cosmology correspond to $x=1/2$.

Once a value for $x$ has been chosen, the step-size of the corresponding
equi-spaced variable $Q$ is determined by another parameter $\delta$, which
appears in shift operators $\exp(-i\delta\hat{P})$. One then replaces the
classical $\dot{a}^2$ in the Friedmann equation by a suitable combination of
exponentials, such as $\delta^{-2}\sin^2(\delta P)$ times some power of $Q$,
and quantizes them to shift operators. (Sometimes, the first replacement step
is done only implicitly.)  Instead of the Wheeler--DeWitt equation, the
classical constraint turns into a difference equation \cite{IsoCosmo} of the
form
\begin{eqnarray} \label{Diff}
&& \frac{2\pi G(1-x)^2}{3\delta^2} \left( 
|(Q-2\delta\hbar)(Q-\delta\hbar)| \psi_{Q-2\delta\hbar}-
 (Q^2+(Q+\delta\hbar)^2)\psi_{Q}\right.\nonumber\\
&&\qquad+\left.
 |(Q+2\delta\hbar)(Q+\delta\hbar)|\psi_{Q+2\delta\hbar}\right)
+ \frac{1}{2}\hat{p}_{\phi}^2\psi_{Q}=0
\end{eqnarray}
for wave functions $\psi_Q(\phi)$.

In (\ref{WdW}) and (\ref{Diff}), the ordering we wrote enjoys special
properties: It makes the system harmonic and free of quantum back-reaction
\cite{BouncePert}. We note that this property is a mathematical feature, which
by no means implies that the ordering is preferred on physical
grounds. However, it is important even for a physical analysis because one can
study more-general orderings by perturbation theory around the harmonic model
\cite{BouncePot,BounceSqueezed,Harmonic} (much as one determines the physics
of interacting quantum field theories by perturbation theory around a free
theory.) The absence of quantum back-reaction makes it convenient to compute
fluctuation energies. Factor orderings other than the one used in (\ref{WdW})
or (\ref{Diff}) imply additional fluctuation terms which one can compute once
the harmonic model is understood.

By solving the Wheeler--DeWitt equation or the difference equation, we obtain
``relational'' wave functions of the form $\psi(a,\phi)$ or $\psi_Q(\phi)$
instead of time-dependent ones. In this situation, there is no clear meaning
of energy eigenvalues where zero-point or fluctuation energies could show
up. Nevertheless, expectation values computed for $\hat{Q}$
at fixed $\phi$ in a solution to the difference equation may be subject to
fluctuation-dependent quantum corrections compared to the classical
relationship between $Q$ and $\phi$. By comparing the behavior of the
expectation value with the classical scale factor, these quantum corrections
can be interpreted as additional energy contributions in an effective
Friedmann equation.

As already mentioned, the ordering matters because re-ordering terms imply
additional quantum corrections which depend explicitly on $\hbar$ and can
compete with fluctuation terms. Several other popular choices exist in loop
quantum cosmology, which follow the same principles of \cite{IsoCosmo},
leading to (\ref{Diff}), but quantize the relevant operators differently. Of
special importance for our purpose will be the treatment of the energy
operator provided by the free, massless scalar, a quantization of the
classical expression (\ref{rhofree}). Corresponding operators have been
defined in different ways and analyzed for instance regarding their
boundedness properties \cite{DensityOp}.

When fluctuation terms are interpreted as energy contributions, their precise
form depends on how one defines a density expectation value. One may, for
instance, define a density operator of the form $\hat{\rho}=\frac{1}{2}
\hat{p}_{\phi}^2\hat{V}^{-2}$ (provided the volume operator $\hat{V}$ is
invertible, which is not always the case and sometimes requires additional
corrections as per \cite{InvScale}). In some suitable state, this operator
then gives rise to an expectation value $\langle\hat{\rho}\rangle$ which one
may take as a measure for the energy density.  Alternatively, and perhaps more
reliably if one considers that canonical quantum gravity provides operators
for matter Hamiltonians but not densities, one may define a measure for the
energy density as $\langle\hat{E}\rangle/\langle\hat{V}\rangle$ with the
energy operator $\hat{E}=\frac{1}{2} \hat{p}_{\phi}^2\hat{V}^{-1}$.

These two expressions differ by fluctuation terms, as shown by an expansion in
moments of the state used. For this purpose, and also for later use, we follow
\cite{EffAc} and introduce the moments
\begin{equation}
 \Delta(V^aP^b\phi^cp_{\phi}^d):= \langle(\hat{V}-\langle\hat{V}\rangle)^a
 (\hat{P}-\langle\hat{P}\rangle)^b (\hat{\phi}-\langle\hat{\phi}\rangle)^c
 (\hat{p}_{\phi}-\langle\hat{p}_{\phi}\rangle)^d\rangle_{\rm Weyl}
\end{equation}
using totally symmetric (Weyl) ordering. (More generally, we should refer to
our $Q$ instead of $V$. For the present example, we assume $x=-1/2$, so that
the volume, up to a constant factor, is one of our basic variables.) For
$a+b+c+d=2$, we have fluctuations and covariance parameters, such as
$\Delta(V^2)=(\Delta V)^2$. Any expression defined as the expectation value of
some Weyl-ordered operator
$\hat{O}=O(\hat{V},\hat{P},\hat{\phi},\hat{p}_{\phi})$ formed from our basic
operators can be expanded in moments by a formal Taylor series
\begin{eqnarray} \label{O}
 \langle \hat{O}\rangle &=& \langle O(\langle\hat{V}\rangle+
 (\hat{V}-\langle\hat{V}\rangle),\langle\hat{P}\rangle+
 (\hat{P}-\langle\hat{P}\rangle), \langle\hat{\phi}\rangle+
 (\hat{\phi}-\langle\hat{\phi}\rangle), \langle\hat{p}_{\phi}\rangle+
 (\hat{p}_{\phi}-\langle\hat{p}_{\phi}\rangle))\\
&=& O(\langle\hat{V}\rangle, \langle\hat{P}\rangle, \langle\hat{\phi}\rangle,
\langle\hat{p}_{\phi}\rangle)+ \sum_{a,b,c,d} \frac{1}{a!b!c!d!}
\frac{\partial^{a+b+c+d} O(\langle\hat{V}\rangle, \langle\hat{P}\rangle,
  \langle\hat{\phi}\rangle, \langle\hat{p}_{\phi}\rangle)}{\partial^a
  \langle\hat{V}\rangle \partial^b \langle\hat{P}\rangle \partial^c
  \langle\hat{\phi}\rangle \partial^d\langle\hat{p}_{\phi}\rangle}
\Delta(V^aP^b\phi^cp_{\phi}^d)\,. \nonumber
\end{eqnarray}
(If $\hat{O}$ is a Weyl-ordered polynomial in basic operators, this expression
is exact. If $\hat{O}$ is not Weyl-ordered, it can be written as a sum of
Weyl-ordered terms some of which have explicit factors of $\hbar$
\cite{Casimir,Search}. For each of them, (\ref{O}) can be used.)  For the
density expressions, we have
\begin{eqnarray} \label{Dens}
 \langle\hat{\rho}\rangle &=& \frac{1}{2}
 \frac{\langle\hat{p}_{\phi}\rangle^2}{\langle\hat{V}\rangle^2}
 \left(1+\frac{\Delta(p_{\phi}^2)}{\langle\hat{p}_{\phi}\rangle^2}+
   3\frac{\Delta(V^2)}{\langle\hat{V}\rangle^2}-
   4\frac{\Delta(Vp_{\phi})}{\langle\hat{p}_{\phi}\rangle\langle\hat{V}\rangle}
+\cdots \right)\\
 \frac{\langle\hat{E}\rangle}{\langle\hat{V}\rangle} &=& \frac{1}{2}
 \frac{\langle\hat{p}_{\phi}\rangle^2}{\langle\hat{V}\rangle^2}
 \left(1+\frac{\Delta(p_{\phi}^2)}{\langle\hat{p}_{\phi}\rangle^2}+
   \frac{\Delta(V^2)}{\langle\hat{V}\rangle^2}-
   2\frac{\Delta(Vp_{\phi})}{\langle\hat{p}_{\phi}\rangle\langle\hat{V}\rangle}
+\cdots \right)
\end{eqnarray}
to which fluctuations and other moments contribute in different ways.

\subsection{Fluctuation energies}

We extend our introductory example of the harmonic oscillator in order to
discuss a general notion of fluctuation energies relevant for quantum
cosmology. Equation (1) can now be recognized as an example of the expansion
(\ref{O}) in terms of moments, which in this case is exact because the
harmonic energy operator is a polynomial in basic operators. For a general
state, the energy expectation value is obtained as the sum of the classical
energy evaluated in expectation values of basic operators, and a fluctuation
term. For an energy eigenstate, only the fluctuation term remains and
determines the energy eigenvalue in this state.

We can derive the zero-point energy in this formalism if we consider the
moments in more detail. Moments, like expectation values, are dynamical and
may change in time, subject to equations of motion. Time derivatives of
expectation values can be derived from the general formula
\begin{equation} \label{dOdt}
 \frac{{\rm d}\langle\hat{O}\rangle}{{\rm d}t} =
 \frac{\langle[\hat{O},\hat{H}]\rangle}{i\hbar} 
\end{equation}
with the Hamiltonian $\hat{H}=\hat{E}$. Moments contain products of
expectation values; their time derivatives can be obtained from (\ref{dOdt})
using the Leibniz rule. For a fluctuation
$\Delta(q^2)=\langle\hat{q}^2\rangle-\langle\hat{q}\rangle^2$, for instance,
we have
\[
 \frac{{\rm d}\Delta(q^2)}{{\rm d}t}=
 \frac{\langle[\hat{q}^2,\hat{H}]\rangle}{i\hbar} -2\langle\hat{q}\rangle
 \frac{\langle[\hat{q},\hat{H}]\rangle}{i\hbar}\,.
\]
The Hamiltonian of the harmonic oscillator implies
\begin{eqnarray}
 \frac{{\rm d}\Delta(x^2)}{{\rm d}t}&=& \frac{2}{m} \Delta(xp)\\
 \frac{{\rm d}\Delta(xp)}{{\rm d}t}&=& \frac{1}{m}\Delta(p^2)-
 m\omega^2\Delta(x^2)\\ 
 \frac{{\rm d}\Delta(p^2)}{{\rm d}t}&=& -2m\omega^2 \Delta(xp)\,.
\end{eqnarray}
Evaluated for a stationary state, with expectation values and moments constant
in time, these equations imply that $\Delta(xp)=0$ and $\Delta(p^2)=
m^2\omega^2\Delta(x^2)$. If we also require that the uncertainty relation be
saturated, as suitable for the harmonic ground state, we obtain $\Delta(x^2)=
\hbar/(2m\omega)$ and the correct zero-point energy.

These considerations show that zero-point energies are just a special case of
fluctuation energies as they follow from a moment expansion (\ref{O}). In this
form, they can be computed also for quantum cosmology, where we can use
(\ref{O}) to expand a constraint operator $\hat{C}$ instead of the
energy. There will then be fluctuation terms from both the gravity and the
matter contribution of the constraint. The former are more sensitive to the
quantization approach used, as well as to factor-ordering ambiguities.

\subsection{Harmonic cosmology}

The computation of fluctuation energies can be done exactly in harmonic
models. Crucial ingredients, realized by the harmonic oscillator as the prime
example, are a polynomial Hamiltonian for the expansion (\ref{O}) to be exact,
with a degree of at most two in canonical variables. The latter property
ensures that the moments obey evolution equations by which they couple only to
other moments of the same order. For more-complicated systems, one can use
perturbation theory provided one can find a harmonic model sufficiently close
to the one of interest. (An example for a large class of such applications is
the low-energy effective action for anharmonic systems
\cite{EffAcQM,EffAc,Karpacz}.)

The harmonic oscillator is not close to all systems studied in quantum
cosmology. But there are substitutes, which one may consider as harmonic
models of cosmology. The procedure of canonical effective equations, based on
(\ref{O}) and (\ref{dOdt}), is easier to perform if one can work with a
Hamiltonian rather than a constraint. We will therefore start with the
technique of deparameterization, allowing one to reformulate constrained
dynamics as formal evolution with respect to one of the degrees of
freedom. After discussing deparameterized harmonic models, we will turn to
additional ingredients required for a direct treatment of constraints.

The models of interest here can easily be deparameterized by considering the
scalar $\phi$ as an evolution parameter. If it is free and massless, the
momentum $p_{\phi}$ is a constant of motion and never becomes
zero. Accordingly, the time derivative ${\rm d}\phi/{\rm d}t=\{\phi,C\}=
p_{\phi}/a^3$ is non-zero, and $\phi(t)$ is a monotonic function. Instead of
$t$, one may therefore use $\phi$ as a unique parameter along dynamical
trajectories.

Equations of motion are then generated canonically by the function
$p_{\phi}(Q,P)$ obtained by solving the constraint for $p_{\phi}$: If we write
the constraint as $C=\frac{1}{2}(p_{\phi}^2-H(Q,P)^2)/a^3=0$, the
Poisson-bracket relationship ${\rm d}O/{\rm d}t=\{O,C\}$ implies ${\rm
  d}O/{\rm d}\phi=({\rm d}t/{\rm d}\phi) \{O,C\}= (a^3/p_{\phi})
\{O,\frac{1}{2}(p_{\phi}^2-H(Q,P)^2)/a^3\}\approx \frac{1}{2} H(Q,P)^{-1}
\{O,-H(Q,P)^2\} =-\{O,H(Q,P)\}$ for any expression independent of $\phi$, up
to terms that vanish when $C=0$.

For the free, massless scalar in a spatially
flat FRW universe, we start with the Hamiltonian constraint expressed in the
canonical pair $(Q,P)$, which can be written as
\begin{equation} \label{CFact}
  \left(\frac{8\pi G(1-x)}{3}Q\right)^{-3/(2(1-x))} \left(-\frac{8\pi G}{3}
    (1-x)^2 Q^2P^2+ \frac{1}{2}p_{\phi}^2\right) =0\,.
\end{equation}
It is sufficient to set the second parenthesis equal to zero (amounting to a
$Q$-dependent choice of the time coordinate), which we do together with a
simple canonical transformation from $(\phi,p_{\phi})$ to
\begin{equation}
 \lambda:=\sqrt{\frac{16\pi G}{3}}(1-x)\phi \quad,\quad p_{\lambda}=
 \sqrt{\frac{3}{16\pi G}} \frac{p_{\phi}}{1-x}
\end{equation}
to absorb some factors.  The $\lambda$-Hamiltonian $p_{\lambda}$ is then a
quadratic function $H(Q,P)=\pm|QP|$, or $\pm\delta^{-1}|Q\sin(\delta P)|$ for
the loop modification. Without loss of generality, we choose the positive sign
in what follows.

\subsubsection{Wheeler--DeWitt model}

In a Wheeler--DeWitt quantization, there are operators for both $\hat{Q}$ and
$\hat{P}$, and we can quantize
$\hat{H}=\widehat{QP}:=\frac{1}{2}(\hat{Q}\hat{P}+\hat{P}\hat{Q})$ in a
symmetric ordering. (For $x=1/2$, (\ref{CFact}) then leads to (\ref{WdW}).) As
indicated in this expression, we drop the absolute value, with the following
justification: The operator $\widehat{QP}$ is preserved by evolution generated
by $|\widehat{QP}|$. Therefore, a state initially supported on the positive
part of the spectral decomposition of $\widehat{QP}$ will always be supported
on this set. For evolution equations (\ref{dOdt}) without the absolute value
in $\hat{H}$, it is then sufficient to ensure initial states to be supported
on the positive part of the spectrum, which can always be achieved by
projection.

With this simplification, $\hat{H}=\hat{p}_{\lambda}$ is a quadratic
polynomial, giving an exact expansion (\ref{O}):
\begin{equation} \label{pphiWdW}
 p_{\lambda}= \langle\hat{Q}\rangle \langle\hat{P}\rangle+ \Delta(QP)
\end{equation}
with a ``fluctuation'' energy $\Delta(QP)$ (which is rather a covariance). In
order to see the meaning of this energy, we transform the deparameterized
equation (\ref{pphiWdW}) back to a constraint, or an effective Friedmann
equation. We compute ${\rm d}\langle\hat{Q}\rangle/{\rm
  d}\lambda=\langle\hat{Q}\rangle$ using (\ref{dOdt}). We transform the
$\lambda$-derivative to a proper-time derivative by ${\rm
  d}\langle\hat{Q}\rangle/{\rm d}t= ({\rm d}\langle\hat{Q}\rangle/{\rm
  d}\lambda) ({\rm d}\lambda/{\rm d}\phi) ({\rm d}\phi/{\rm d}t)$. Finally,
instead of $\langle\hat{Q}\rangle$ we introduce the effective scale factor
\begin{equation}
 a_{\rm eff}:= \left(\frac{8\pi
     G}{3}(1-x)\langle\hat{Q}\rangle\right)^{1/(2(1-x))}
\end{equation}
and write an effective Friedmann equation
\begin{equation}
 \left(\frac{\dot{a}_{\rm eff}}{a_{\rm eff}}\right)^2 = \left(\frac{1}{2(1-x)}
 \frac{\dot{\langle\hat{Q}\rangle}}{\langle\hat{Q}\rangle}\right)^2 =
 \frac{4\pi G}{3} \frac{p_{\phi}^2}{a_{\rm eff}^6}\,.
\end{equation}
This effective Friedmann equation does not differ from the classical Friedmann
equation for a free, massless scalar. The only implication of the fluctuation
energy in this case is a shift from the classical $p_{\phi}$ to
$p_{\phi}+\sqrt{16\pi G/3} (1-x) \Delta(QP)$. Since $\Delta(QP)$ is constant
under evolution generated by the effective Hamiltonian (\ref{pphiWdW}), a
constant shift of the constant of motion $p_{\phi}$ does not change the form
of the dynamical equation.

For $x=-1/2$, in which case $-P={\cal H}$ is the classical Hubble parameter,
one could try to derive an effective Friedmann equation in a different way, by
defining an effective Hubble parameter ${\cal H}_{\rm
  eff}:=-\langle\hat{P}\rangle$. Starting with (\ref{pphiWdW}), we could then
write
\begin{equation}
 {\cal H}_{\rm eff}^2\left(1+\frac{\Delta(V{\cal H})}{a_{\rm eff}^3{\cal
       H}_{\rm eff}}\right)^2= 
 \frac{4\pi G}{3}\frac{p_{\phi}^2}{a_{\rm eff}^6}
\end{equation}
which does have a moment term. However, even if one uses the fact that
$\Delta(V{\cal H})$ and $p_{\phi}$ are constant, this equation is not a closed
effective equation because ${\cal H}_{\rm eff}$ or $\langle\hat{P}\rangle$ is
in general independent of $a_{\rm eff}$ or $\langle\hat{Q}\rangle$.


\subsubsection{Loop model}
\label{s:Eff}

The situation is more interesting for the constraints of loop quantum
cosmology. These expressions are not polynomial in $(Q,P)$, but they can be
made so if one transforms to non-canonical variables $(Q,J)$ with
$J:=Q\exp(i\delta P)$. Our new variables can still be considered as basic ones
because they form a closed algebra
\begin{equation}
 \{Q,J\}= i\delta J\quad,\quad \{Q,J^*\}= -i\delta J^*\quad,\quad \{J,J^*\}=
 2i\delta Q\,.
\end{equation}
We quantize this basic algebra to
\begin{equation}
 [\hat{Q},\hat{J}]= -\delta\hbar \hat{J}\quad,\quad
 [\hat{Q},\hat{J}^{\dagger}]= \delta\hbar\hat{J}^{\dagger}\quad,\quad
 [\hat{J},\hat{J}^{\dagger}]= -2\delta\hbar\hat{Q}\,.
\end{equation}
These commutators can be shown to follow from an ordering
$\hat{J}=\hat{Q}\widehat{\exp(i\delta P)}$ up to an inconsequential shift of
$\hat{Q}$ by $\hbar/2$ \cite{BouncePert,BounceCohStates}.

Instead of $QP$, the classical $\lambda$-Hamiltonian is then $H(Q,J)=
\delta^{-1} {\rm Im}J$, linear in our new variables. There is therefore no
fluctuation contribution to $\langle\hat{H}\rangle=\delta^{-1}{\rm
  Im}\langle\hat{J}\rangle$. For this operator, we can recognize the ordering
in (\ref{Diff}) by comparing it with
$\hat{p}_{\lambda}^2=-\frac{1}{4}\delta^{-2}
(\hat{J}^2-\hat{J}\hat{J}^{\dagger}-
\hat{J}^{\dagger}\hat{J}+(\hat{J}^{\dagger})^2)$ acting on eigenstates of
$\hat{Q}$.  

With a closed commutator algebra of basic operators and a linear Hamiltonian,
there is no quantum back-reaction of moments coupling dynamically to
expectation values.  However, there is a quadratic relationship between the
variables, $|J|^2-Q^2=0$, in order to ensure that $P$ contained in $J$ is
real. (It implies a Casimir constraint in the sense of \cite{Casimir}; see
also App.~\ref{a:Casimir}.)  Upon quantization, this reality condition takes
the form
\begin{equation} \label{Reality}
 |\langle\hat{J}\rangle|^2- \langle\hat{Q}\rangle^2=
 \Delta(Q^2)-\Delta(J\bar{J})\,.
\end{equation}
If one rewrites the quantum Hamiltonian as an effective Friedmann equation,
one must express ${\rm Re}\langle\hat{J}\rangle$ in terms of
$\langle\hat{H}\rangle$, in which process one uses
(\ref{Reality}). Fluctuation energies are thereby obtained in an indirect
way. We will now show more details, but first note that moments can play
crucial dynamical roles even in models that do not have quantum
back-reaction. The absence of quantum back-reaction implies that equations of
motion for expectation values do not couple to moments of a state, as realized
in (\ref{dQdl}) below. But if there are additional constraints, such as
(\ref{Reality}), equations derived from those for basic expectation values may
include moments; see (\ref{EffFried}) below.

From our Hamiltonian linear in $\hat{J}$ we obtain an equation of motion
\begin{equation} \label{dQdl}
  \frac{{\rm d}\langle\hat{Q}\rangle}{{\rm d}\lambda}= -\frac{1}{2\delta\hbar}
  \langle[\hat{Q},\hat{J}-\hat{J}^{\dagger}]\rangle= \frac{1}{2}
  \langle\hat{J}+\hat{J}^{\dagger}\rangle  = {\rm
    Re}\langle\hat{J}\rangle\,.
\end{equation}
We can use this equation to compute ${\rm d}\langle\hat{Q}\rangle/{\rm d}t=
({\rm d}\phi/{\rm d}t)({\rm d}\lambda/{\rm d}\phi) ({\rm
  d}\langle\hat{Q}\rangle/{\rm d}\lambda)$ with ${\rm d}\phi/{\rm
  d}t=p_{\phi}/(\frac{8}{3}\pi G(1-x)\langle\hat{Q}\rangle)^{3/2(1-x)}$ as
before. But first, we use the reality condition (\ref{Reality}) in order to
express ${\rm Re}\langle\hat{J}\rangle$ in terms of ${\rm
  Im}\langle\hat{J}\rangle$ and the fluctuation parameter
\begin{equation} \label{epsilon}
 \epsilon:= \frac{\Delta(Q^2)-\Delta(J\bar{J})}{\langle\hat{Q}\rangle^2}\,.
\end{equation}
We obtain
\begin{eqnarray}
 \frac{1}{\langle\hat{Q}\rangle^2} \left(\frac{{\rm
       d}\langle\hat{Q}\rangle}{{\rm d}\phi}\right)^2 &=& \frac{16\pi G}{3}
 (1-x)^2 \frac{({\rm
     Re}\langle\hat{J}\rangle)^2}{\langle\hat{Q}\rangle^2} \label{dQdphi} \\
&=& \frac{16\pi G}{3}(1-x)^2 \frac{\langle\hat{Q}\rangle^2- ({\rm
    Im}\langle\hat{J}\rangle)^2+
\epsilon\langle\hat{Q}\rangle^2}{\langle\hat{Q}\rangle^2}\,.\nonumber
\end{eqnarray}
We then use
\[
 {\rm Im}\langle\hat{J}\rangle= \delta
\langle\hat{H}\rangle= \sqrt{\frac{3}{16\pi G}} \frac{\delta
   p_{\phi}}{1-x}
\]
and
\[
 \frac{({\rm Im}\langle\hat{J}\rangle)^2}{\langle\hat{Q}\rangle^2}=
 \frac{3}{16\pi G} \frac{\delta^2p_{\phi}^2}{(1-x)^2 \langle\hat{Q}\rangle^2}=
 \frac{8\pi G}{3} \delta^2 \frac{p_{\phi}^2}{2a_{\rm eff}^6} \left(\frac{8\pi
     G}{3} (1-x) \langle\hat{Q}\rangle\right)^{(1+2x)/(1-x)}\,.
\]
Therefore,
\begin{equation}
\frac{1}{\langle\hat{Q}\rangle^2} \left(\frac{{\rm
       d}\langle\hat{Q}\rangle}{{\rm d}\phi}\right)^2
=\frac{16\pi G}{3}(1-x)^2 \left(1+\epsilon-
  \frac{\rho_{\rm free}}{\rho_{\rm QG}}\right)
\end{equation}
with
\begin{equation} \label{rhoQG}
 \rho_{\rm QG} := \frac{3}{8\pi G\delta^2 (\frac{8}{3}\pi
G(1-x)\langle\hat{Q}\rangle)^{(1+2x)/(1-x)}}\,.
\end{equation}
(The energy density of the free, massless scalar is $\rho_{\rm free}=
\frac{1}{2}p_{\phi}^2 (\frac{8}{3}\pi
G(1-x)\langle\hat{Q}\rangle)^{-3/(1-x)}$.)  We finally write the equation for
the proper-time derivative of $a_{\rm eff}$ in the form of an effective
Friedmann equation
\begin{eqnarray} 
\left(\frac{\dot{a}_{\rm eff}}{a_{\rm
        eff}}\right)^2&=& \left(\frac{1}{2(1-x)} \frac{{\rm d}\phi}{{\rm d}t}
    \frac{{\rm d}\langle\hat{Q}\rangle/{\rm d}\phi}{\langle\hat{Q}\rangle}
  \right)^2 \nonumber\\
&=& \frac{8\pi G}{3} \rho_{\rm free}
  \left(1+\epsilon -\frac{\rho_{\rm free}}{\rho_{\rm QG}}\right)= \frac{8\pi
    G}{3} \left(\rho_{\rm free}+\rho_{\rm free}\epsilon- \frac{\rho_{\rm
        free}^2}{\rho_{\rm QG}}\right)\,.\label{EffFried} 
\end{eqnarray}

In (\ref{EffFried}), the parameter $\epsilon$, multiplied with $\rho_{\rm
  free}$, plays the role of a fluctuation energy and affects the dynamics. It
may seem surprising that $\rho_{\rm free}\epsilon$ is added to the energy
density, rather than just $\epsilon$.  However, using $\rho_{\rm free}=
\frac{1}{2}({\rm d}\phi/{\rm d}t)^2$, we observe that there are two factors of
${\rm d}\phi/{\rm d}t$ in $\rho_{\rm free}\epsilon$ which transform the
fluctuation energy $\epsilon$ (or rather $\frac{1}{2}\epsilon$) from the
$\phi$-frame to the $t$-frame, according to the tensor-transformation law
$\rho=T_{00}= ({\rm d}t'/{\rm d}t)^2T_{0'0'}= ({\rm d}t'/{\rm d}t)^2\rho'$ in
isotropic models, in which only the time coordinate is being changed.

This observation highlights one of the coincidences realized for the free,
massless scalar source, for which ${\rm d}\phi/{\rm d}t=\sqrt{2\rho_{\rm
    free}}$. If there is mass term or a potential $W(\phi)$, it is not known
how to generalize (\ref{EffFried}) except for perturbative derivations for
small potential \cite{BounceSqueezed}. But it is clear that the factor of
$\rho_{\rm free}$ multiplying the parenthesis in (\ref{EffFried}) plays
different roles for the three terms: The classical Friedmann equation requires
the total energy density $\rho=\rho_{\rm free}+W(\phi)$, while $\epsilon$
should be multiplied only with $\rho_{\rm free}$ to provide the correct
transformation of frames. For this reason, as well as the presence of quantum
back-reaction, it is not easy to generalize (\ref{EffFried}) to a scalar with
mass or self-interactions.

\section{Applications}
\label{s:App}

Our derivation of fluctuation energies in the preceding section has clarified
the physical meaning of moment-dependent terms in effective equations of
harmonic models. The formal part of our calculations, however, was not new
compared with previous treatments of effective equations. In this section, we
use our results for two novel applications.

\subsection{Instability of initial-value formulations in Euclidean regimes}

The effective Friedmann equation (\ref{EffFried}) shows that $a_{\rm eff}(t)$
has a turning point when the energy density $\rho_{\rm free}$ reaches the
value $\rho_{\rm QG}(1+\epsilon)$. This turning point has been interpreted as
a bounce, but since consistent inhomogeneous extensions of the background
model show that space-time turns into a quantum version of 4-dimensional
Euclidean space at high density, there is no deterministic evolution and the
bounce picture is incorrect. Instead of using an initial-value formulation,
the Euclidean phase with elliptic mode equations requires a boundary-value
problem including the $t$-direction.

If one were to use an initial-value problem throughout the Euclidean phase,
one could still find solutions to the partial differential equations for
inhomogeneities. However, these solutions are not stable and depend
sensitively on the initial values one selects. (Instead of oscillating Fourier
terms $\exp(\pm i\omega t)$ one has exponential ones $\exp(\pm\omega t)$.) The
Euclidean phase can be shown to occupy a small Planckian $t$-range for the
harmonic model when fluctuations are small. Instabilities therefore do not
make sub-Planckian fields with $\omega\ll t_{\rm P}^{-1}$ grow much and may be
assumed harmless.\footnote{This possibility has been pointed out by Jaume
  Garriga.} In this section, we show that the situation changes for large
fluctuations, which one should expect in a generic quantum regime likely to be
realized at high density. (Our arguments indicate that a potential will have
the same effect, although in this case it is more difficult to generalize the
effective Friedmann equation.)

Consistent inhomogeneous extensions have been derived for the modified
Friedmann equation (\ref{ModFried}), which follows from a modified constraint
\begin{equation} \label{Cmod} 
C_{\rm mod}=-\frac{3}{8\pi G}
  V\frac{\sin^2(\delta P)}{\delta^2}+ \frac{p_{\phi}^2}{2V}=0
\end{equation}
assuming $x=-1/2$ and referring to the pair $(V,P)$ with $\{V,P\}=4\pi
G$. Indeed, if we compute $\dot{V}=\{V,C_{\rm mod}\}= -3V\sin(2\delta
P)/2\delta$, we have ${\cal H}^2=(\frac{1}{3}\dot{V}/V)^2=
\delta^{-2}(\sin^2(\delta P)-\sin^4(\delta P))= (8\pi G/3) \rho_{\rm free}
(1-\rho_{\rm free}/\rho_{\rm QG})$. Alternatively, we can write this equation
as $\sin^2(\delta P)=\rho_{\rm free}/\rho_{\rm QG}$. For a background dynamics
subject to this modification, consistent mode equations have the speed
$\beta=\cos(2\delta P)=1-2\rho_{\rm free}/\rho_{\rm QG}$
\cite{ScalarHol,ScalarHolInv}. More generally, if the term ${\cal H}^2$ in the
classical Friedmann equation is replaced by some function $f(V,P)$ with ${\cal
  H}^2$ as the limit for $\delta\to0$, the speed of modes in consistent
inhomogeneous models is given by
\begin{equation} \label{beta}
\beta=\frac{1}{2}\partial^2f/\partial P^2\,.
\end{equation}
This general form, derived for spherically symmetric models
\cite{JR,HigherSpatial}, is consistent with the results of
\cite{ScalarHol,ScalarHolInv}. When $\beta<0$, mode equations become elliptic
and the space-time signature turns Euclidean. In cosmological models based on
(\ref{ModFried}), the density at the transition point is half the maximum
density, $\frac{1}{2}\rho_{\rm QG}$.

Consistent versions of inhomogeneous equations have not yet been derived in
the presence of moment terms and quantum back-reaction, and we cannot easily
extend these conclusions about signature change to the effective Friedmann
equation (\ref{EffFried}) when the fluctuation energy $\epsilon$ is
large. Fortunately, however, the general form of the relationship (\ref{beta})
allows us to estimate the behavior with just a few reasonable assumptions.

For a consistent set of equations, the moment dependence of background and
mode equations is likely to be restricted. Without knowing the precise
dependence, we only assume that the modification function in the constraint
\begin{equation}
 C_{\rm mod}=-\frac{3}{8\pi G} V f(V,P,\Delta(\cdot)) + V\rho_{\rm free}=0
\end{equation}
is now allowed to depend also on moments $\Delta(\cdot)$ of the pair
$(V,P)$. If the dependence of $f$ on $V$ is not very strong, we still have a
monotonic function $P(t)$ because
\begin{equation}
 \frac{{\rm d}P}{{\rm d}t}=\{P,C_{\rm mod}\}= \frac{3}{2}
 \left(f+V\frac{\partial f}{\partial V}\right)+2\pi
 G\frac{p_{\phi}^2}{V^2}\approx 2f+\frac{3}{2}V\frac{\partial f}{\partial V}>0
\end{equation} 
as long as $\partial f/\partial V$ is sufficiently small. (We have $f=(8\pi
G/3) \rho_{\rm free}>0$ when $C_{\rm mod}=0$.) With (\ref{beta}), $\beta$ is
therefore negative when $\partial f/\partial P$ decreases in time.

The change of $\partial f/\partial P$ in time is related to the behavior of
the effective Hubble parameter: we have $\dot{a}_{\rm eff}/a_{\rm eff}=
\frac{1}{3}\dot{V}/V= \{V,C_{\rm mod}\}/3V= -\frac{1}{2} \partial f/\partial
P$. We can then write
\begin{equation} \label{betaa}
 \beta=\frac{1}{2}\partial^2f/\partial^2P=- {\rm
  d}(\dot{a}_{\rm eff}/a_{\rm eff})/{\rm d}P\,.
\end{equation}
All we need to do to determine the density range of the Euclidean phase is to
discuss the behavior of the effective Hubble parameter in relation to the
energy density, as given by the effective Friedmann equation.

In a collapse phase, the energy density increases and approaches $\rho_{\rm
  QG}(1+\epsilon)$. For sufficiently small densities, $-\dot{a}_{\rm
  eff}/a_{\rm eff}$ is positive and increases until $\delta^2(\dot{a}_{\rm
  eff}/a_{\rm eff})^2=(\rho_{\rm free}/\rho_{\rm QG})(1+\epsilon- \rho_{\rm
  free}/\rho_{\rm QG})$ reaches a maximum as a function of $\rho_{\rm
  free}/\rho_{\rm QG}$. From then on, $-\dot{a}_{\rm eff}/a_{\rm eff}$
decreases with $P$ and $\beta$ becomes negative according to
(\ref{betaa}). The maximum is reached when $\rho_{\rm
  free}=\frac{1}{2}\rho_{\rm QG}(1+\epsilon)$, that is at half the maximum
density where $\dot{a}_{\rm eff}/a_{\rm eff}=0$. When the density is between
$\frac{1}{2}\rho_{\rm QG}(1+\epsilon)$ and $\rho_{\rm QG}(1+\epsilon)$, the
signature is of Euclidean type.

The harmonic model provides solutions $\langle\hat{V}\rangle(\phi)\propto
\cosh(\phi)$ \cite{BouncePert}. (See also App.~\ref{a:ScalarTime}.) For the
free density $\frac{1}{2} p_{\phi}^2/\langle\hat{V}\rangle^2$ to change from
$\frac{1}{2}\rho_{\rm QG}(1+\epsilon)$ to $\rho_{\rm QG}(1+\epsilon)$ (with
constant $p_{\phi}$), we need $\cosh(\phi)$ to change by a factor of the order
one. Therefore, $t$ changes by an amount $\Delta t$ which is a number of order
one times the (nearly constant) value of ${\rm d}t/{\rm
  d}\phi=1/\sqrt{2\rho_{\rm free}}$ in this small $\phi$-interval. In the
given density range, we have $\rho_{\rm free}\sim \rho_{\rm
  QG}(1+\epsilon)$. If fluctuations are significant and $\epsilon<0$,
$\rho_{\rm free}$ can be well below the Planck density even if $\rho_{\rm
  QG}\sim \rho_{\rm P}$. (In the next section we will show that $\epsilon$ is
negative for a Gaussian in $V$.) Accordingly, the $t$-range of the Euclidean
phase is much larger than the $\phi$-range, which is of order one. Quantum
fluctuations can enlarge the size of the Euclidean phase, so that
instabilities of an initial-value formulation are relevant not only for
trans-Planckian modes. Only a boundary-value formulation for elliptic
equations can avoid these instabilities.

\subsection{Comparisons with numerical results}

If one solves differential or difference equations numerically, one must
assume an initial wave function. Unfortunately, quantum cosmology does not
give rise to a strongly restricted class of states. Gaussians in some
variables are usually justified in near-vacuum considerations of perturbative
field theories, just because the free vacuum happens to be Gaussian. Quantum
cosmology, with its unbounded-from-below gravitational contribution to the
Hamiltonian constraint, does not imply a clear ground state, let alone a
near-Gaussian one. And although Gaussians provide nice semiclassical states, a
general semiclassical regime may require a larger class of states (perhaps
even mixed ones). In models with relevant quantum back-reaction or fluctuation
energies, the form of the state matters because it determines the moments.

Nevertheless, provided one interprets them carefully enough in the light of
quantization ambiguities and state choices, numerical solutions can provide
valuable insights. In this section, we discuss several examples of fluctuation
effects which can be derived easily from effective equations and be compared
with existing numerical results.

\subsubsection{Fluctuation energy}
\label{s:FluctEn}

Our expression for the fluctuation energy in (\ref{EffFried}), given by
(\ref{epsilon}), is valid for any state. We illustrate its effects more
explicitly by computing the value for two classes of states, both Gaussian but
with respect to different canonical variables.

\paragraph{Gaussian in the volume.}

Given the expression of $\epsilon$ in terms of $(V,P)$-moments, one can rather
easily compute it for a Gaussian state in $Q$, with wave function
\begin{equation}
 \psi(Q)= \frac{1}{(2\pi)^{1/4}\sqrt{\sigma}} \exp\left(-
 \frac{(Q-\bar{Q})^2}{4\sigma^2}\right) \exp(i\hbar^{-1}\bar{P}Q)\,.
\end{equation}
(For all wave functions, we refer to standard $L^2$-Hilbert spaces. Since we
have deparameterized by $\phi$ or $\lambda$, these are physical Hilbert
spaces.)  The $Q$-fluctuation $\Delta Q=\sigma$ and the expectation values
$\bar{Q}$ and $\bar{P}$ of $\hat{Q}$ and $\hat{P}$ are well-known.

Using $\hat{J}=\hat{Q}\widehat{\exp(i\delta P)}$, with the exponential acting
as a shift operator on $\psi(Q)$, we compute
\begin{equation}
 \langle\hat{J}\rangle= (\bar{Q}-{\textstyle\frac{1}{2}} \delta\hbar)
 \exp(-{\textstyle\frac{1}{8}}
 \sigma^{-2}\delta^2\hbar^2+i\bar{P}\delta) 
\end{equation}
and
\begin{equation}
 \Delta(J\bar{J})=
 \sigma^2+\bar{Q}(\bar{Q}-\delta\hbar)
\left(1-\exp(-\delta^2\hbar^2/4\sigma^2)\right)+
 \frac{1}{2}\delta^2\hbar^2
 \left(1-\frac{1}{2}\exp(-\delta^2\hbar^2/4\sigma^2)\right)\,.
\end{equation}
For $2\sigma\gg\delta\hbar$ (a $Q$-fluctuation much larger than the discrete
spacing of $Q$), we can expand the exponentials and obtain
\begin{equation} \label{epsilonQ}
 \epsilon=-\frac{1}{4}\frac{\delta^2\hbar^2}{\sigma^2}=
-\frac{1}{4}\frac{\delta^2\hbar^2}{(\Delta
  Q)^2}=-\frac{4\pi^2\delta^2\ell_{\rm P}^4}{(\Delta V)^2}\,,
\end{equation}
the last part for $x=-1/2$.  

\begin{figure}
\begin{center}
\includegraphics[width=9.5cm]{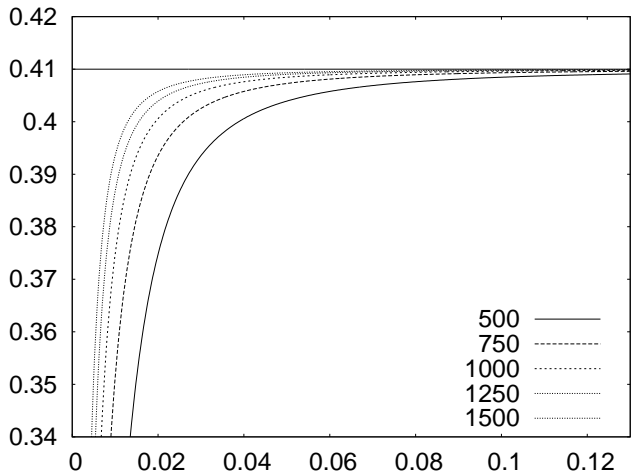}
\caption{The maximal effective density (\ref{EffDensV}) as a function of
  $\Delta V/\langle\hat{V}\rangle$, for different values of $p_{\phi}$. The
  parameter $\rho_{\rm QG}\approx 0.41$ (in units with $G=1=\hbar$)
  has been chosen to be close to the one used in \cite{NumBounce}. (This
  choice amounts to $\delta=2\sqrt{\pi} \sqrt[4]{3} \gamma^{3/2}$ with the
  Barbero--Immirzi parameter $\gamma=0.238$ \cite{Gamma,Gamma2}, as it follows
  from a comparison of the general (and non-unique) step-size $2\delta$ in
  (\ref{Diff}) with the specific choice made in \cite{NumBounce}.) This plot
  is to be compared with Fig.~15 of \cite{NumBounce}. \label{Fig:GaussianV}}
\end{center}
\end{figure}

The fluctuation energy for a $Q$-Gaussian is inversely proportional to the
squared $Q$-fluctuation and grows for small $\Delta Q$, and it is
negative. For a comparison with \cite{NumBounce} (using $x=-1/2$), it is
useful to replace the dependence on $\Delta V$ only by a dependence on the two
parameters $\Delta V/\langle\hat{V}\rangle$ and $p_{\phi}$, writing
\begin{equation}
 \epsilon= -\frac{3\pi G\hbar^2}{p_{\phi}^2}
 \left(\frac{\langle\hat{V}\rangle}{\Delta V}\right)^2 \frac{\rho_{\rm
     free}}{\rho_{\rm QG}}\,.
\end{equation}
If $\Delta V/\langle\hat{V}\rangle$ and $p_{\phi}$ are treated as independent
variables, $\epsilon$ depends on $\rho_{\rm free}$. Our previous equation for
the maximum effective density, $\rho_{\rm free}^{\rm max}= \rho_{\rm
  QG}(1+\epsilon)$, can then be solved for
\begin{equation} \label{EffDensV}
 \rho_{\rm free}^{\rm max} = \frac{\rho_{\rm QG}}{1+\frac{3\pi
     G\hbar^2}{p_{\phi}^2} \left(\frac{\langle\hat{V}\rangle}{\Delta
       V}\right)^2}\,.
\end{equation}
See Fig.~\ref{Fig:GaussianV}.

Slightly more generally, we can allow for correlations of the pair $(Q,P)$,
amounting to a fully squeezed state with wave function
\begin{equation}
 \psi(Q)= \frac{1}{(2\pi)^{1/4}\sqrt{\sigma}} \exp\left(-
 \frac{(Q-\bar{Q})^2}{4\sigma^2}(1-2i\kappa/\hbar)\right)
\exp(i\hbar^{-1}\bar{P}Q) 
\end{equation}
with the covariance $\kappa=\Delta(QP)$. We still have $\Delta Q=\sigma$, but
$\kappa$ contributes to the momentum fluctuation $\Delta
P=\frac{1}{2}\hbar\sqrt{1+4\kappa^2/\hbar^2}/\sigma$.

Proceeding as before, we now have 
\begin{eqnarray*}
 \Delta(J\bar{J})&=&
 \sigma^2+\bar{Q}(\bar{Q}-\delta\hbar)
\left(1-\exp(-\delta^2\hbar^2(1+4\kappa^2/\hbar^2)/4\sigma^2)\right)\\
&&+
 \frac{1}{2}\delta^2\hbar^2
 \left(1-\frac{1}{2}(1+4\kappa^2/\hbar^2)
\exp(-\delta^2\hbar^2(1+4\kappa^2/\hbar^2)/4\sigma^2)\right)\,. 
\end{eqnarray*}
The dominant contribution to $\epsilon$ receives an additional factor of
$1+4\kappa^2/\hbar^2$. As this factor is always positive, correlations do not
change the fact that a Gaussian in the volume does not increase the maximum
density beyond $\rho_{\rm QG}$. The graph of $\rho_{\rm free}^{\rm
  max}$ is as in Fig.~\ref{Fig:GaussianV} but with values
$p_{\phi}/\sqrt{1+4\kappa^2/\hbar^2}$ instead of $p_{\phi}$.

\paragraph{Gaussian in the scalar.}

In \cite{APSII} and \cite{NumBounce}, an alternative wave function has been
used as an initial state in a regime in which the Wheeler--DeWitt equation is
valid, which amounts to a Gaussian as well but in $(\phi,p_{\phi})$ rather
than $(Q,P)$. It has the form
\begin{equation}\label{phiGaussian}
 \psi(Q)= \sqrt{\frac{\sigma/2}{\sqrt{2\pi}|Q|}} \exp\left(-\frac{1}{4}\sigma^2
   (\log |Q/\bar{Q}|)^2\right) \exp(i\bar{p}_{\lambda}
 \log(|Q/\bar{Q}|)/\hbar)
\end{equation}
in a slightly modified notation. If (\ref{phiGaussian}) is used as a state in
the loop-quantized model, the inner product is initially defined by summation
over a discrete subset of all $Q$, but for states spread more widely than
$\delta\hbar$ the summation is well approximated by an integral.

In a quantum model deparameterized by $\phi$ or $\lambda$ it is, in general,
not meaningful to speak of a Gaussian in $(\lambda,p_{\lambda})$. Moreover,
there are no operators for $\hat{\lambda}$ and $\hat{p}_{\lambda}$, and
correspondingly no expectation values and moments for these
variables. However, in an initial-state regime in which the Wheeler--Dewitt
equation is valid and the evolved state remains semiclassical, a
$\lambda$-Gaussian $\psi(Q)$ may be defined in
$\lambda(Q)=\log|Q/\bar{Q}|+\bar{\lambda}$, the classical solution for
$\lambda$. This definition leads to (\ref{phiGaussian}). As for operators, one
can use the constraint equation to replace
$\hat{p}_{\lambda}\psi=-i\hbar\partial\psi/\partial\lambda$ by
$\widehat{QP}\psi$, and $\hat{\lambda}$ as a multiplication operator with the
classical solution $\lambda(Q)=\log|Q/\bar{Q}|+\bar{\lambda}$. With these
prescriptions, we obtain
\begin{equation} \label{Deltalambda}
 \langle\hat{\lambda}\rangle=\bar{\lambda}\quad,\quad
 \langle\hat{p}_{\lambda}\rangle=\bar{p}_{\lambda} \quad,\quad
 \Delta\lambda=\frac{1}{\sigma}\quad,\quad \Delta p_{\lambda}=
 \frac{1}{2}\hbar \sigma
\end{equation}
for a wave function (\ref{phiGaussian}).

Moreover, in the standard way of a $Q$-representation of wave functions we
compute the expectation value
\begin{equation}
  \langle\hat{Q}\rangle = \frac{\sigma}{\sqrt{2\pi}} \int_0^{\infty} \exp\left(-
    \frac{\sigma^2}{2} (\log(Q/\bar{Q}))^2\right){\rm d}Q= \bar{Q}
  \exp(1/2\sigma^2)\,.
\end{equation}
(The integration can easily be performed after substituting
$\lambda=\log(Q/\bar{Q})$, so that also here we are formally writing the wave
function as a Gaussian in the scalar.)  A similar calculation gives
\begin{equation} \label{DeltaQ}
 \Delta(Q^2)= \bar{Q}^2 \left(e^{2/\sigma^2}- e^{1/\sigma^2}\right)=
   \langle\hat{Q}\rangle^2 (e^{1/\sigma^2}-1)\,.
\end{equation}

In order to compute expectation values containing shift operators, such as
$\langle\hat{J}\rangle$, we may expand in $\delta\hbar/Q$ before integrating,
which is reasonable in the same regime in which the inner product can be
written as an integral. However, we must be careful because the momentum term
$\exp(i\hbar^{-1}\bar{p}_{\lambda} \log|(Q+\delta\hbar)/\bar{Q}|)$, after the
logarithm is expanded in $\delta\hbar/Q$, gives rise to terms of the order
$\delta\bar{p}_{\lambda}/Q$ which are not small near the bounce
regime. Nevertheless, it may be of interest to expand in
$\delta\bar{p}_{\lambda}/Q$ for analytic integrations in the form of
polynomials times Gaussians. When applied to an initial state, the volume
spread should therefore be sufficiently small to keep the support of the wave
function away from $Q\sim\delta\bar{p}_{\lambda}$.

In this way, we find
\begin{equation}
 \langle\hat{J}\rangle = \langle\hat{Q}\rangle-\frac{1}{2} \delta\hbar+
 i\delta \bar{p}_{\lambda}
\end{equation}
(note that ${\rm Im}\langle\hat{J}\rangle=\delta \bar{p}_{\lambda}$,
consistent with $\langle\hat{p}_{\lambda}\rangle=\bar{p}_{\lambda}$) and the
final moment
\begin{equation}
 \Delta(J\bar{J})= \langle\hat{Q}^2\rangle - \delta \hbar
 \langle\hat{Q}\rangle- |\langle\hat{J}\rangle|^2\,.
\end{equation}
The fluctuation energy in this case is
\begin{equation} \label{EpsilonCancel}
 \epsilon= \frac{\delta^2 \bar{p}_{\lambda}^2}{\langle\hat{Q}\rangle^2}= \frac{4\pi
   G}{3} \frac{\delta^2 p_{\phi}^2}{\langle\hat{V}\rangle^2}= \frac{\rho_{\rm
     free}}{\rho_{\rm QG}}\,,
\end{equation}
the latter relations for $x=-1/2$. Surprisingly, the fluctuation energy (which
is now positive) cancels the term produced by holonomy modifications in the
Friedmann equation. (The next-order contribution $4\pi^2\delta^2\ell_{\rm
  P}^2/\langle\hat{V}\rangle^2$ is very small.) Moment terms are thereby shown
to be able to rival holonomy modifications of loop quantum cosmology. If such
a state were realized for an extended period of evolution at high density (as
opposed to initially as assumed in \cite{APSII,NumBounce}), the volume
expectation value could avoid a bounce. But again, our approximation of the
integrations in $\Delta(J\bar{J})$, especially in ${\rm
  Re}\langle\hat{J}\rangle$, is not expected to be good at high density.

Also here, we can repeat our calculations for fully squeezed states with wave
function 
\begin{equation}\label{phiGaussianC}
 \psi(Q)= \sqrt{\frac{\sigma/2}{\sqrt{2\pi}|Q|}} \exp\left(-\frac{1}{4}\sigma^2
   (\log |Q/\bar{Q}|)^2(1-2i\kappa/\hbar)\right) \exp(i\bar{p}_{\lambda}
 \log(|Q/\bar{Q}|)/\hbar)\,.
\end{equation}
In this case, we find that the fluctuation energy does not depend on $\kappa$.

\subsubsection{Effective constraints and relations between moments}

So far, we have mainly considered deparameterized equations. After
deparameterization, the ``time'' part of the system, given here by
$(\phi,p_{\phi})$ or $(\lambda,p_{\lambda})$, is not fully quantized, and
information about moments containing one or more of these variables is
partially lost. We have to go back to the original constrained system in order
to retrieve this information, while making sure that the constraints are
satisfied for physical states. The formalism of effective constraints,
developed in \cite{EffCons,EffConsRel}, is useful for this task.

A constraint operator $\hat{C}$, such as
\begin{equation} \label{C2}
\hat{C}=\delta^2\hat{p}_{\lambda}^2-(\widehat{Q\sin(\delta P)})^2
\end{equation}
as used for (\ref{Diff}), gives rise to an effective constraint
$\langle\hat{C}\rangle$ which can be expanded in expectation values and
moments just like an effective Hamiltonian, using (\ref{O}). However, solving
$\langle\hat{C}\rangle=0$ is not sufficient because a vanishing expectation
value of $\hat{C}$ in some state does not imply that the state is annihilated
by $\hat{C}$. As shown in \cite{EffCons,EffConsRel}, a complete and consistent
(first-class) constrained system is obtained if one accompanies
$\langle\hat{C}\rangle$ by infinitely many effective constraints $C_{\rm
  pol}:=\langle\widehat{\rm pol}\hat{C}\rangle$ for polynomials $\widehat{\rm
  pol}$ in $\hat{O}-\langle\hat{O}\rangle$ for all basic operators
$\hat{O}$. To finite order in the moments, a finite number of effective
constraints is sufficient.

As basic operators appropriate for (\ref{C2}), we choose
$(\hat{Q},\widehat{\sin(\delta P)}, \widehat{\cos(\delta P)})$.  There will
then be moments involving powers of all three operators. However, they are not
all independent if we impose the constraint
\begin{equation} \label{T}
 \hat{T}\psi:= \left(\widehat{\sin(\delta P)}^2+ \widehat{\cos(\delta
     P)}^2-1\right)\psi=0
\end{equation}
so that $\langle\widehat{\exp(i\delta P)}\rangle= \langle\widehat{\cos(\delta
  P)}\rangle+i \langle\widehat{\sin(\delta P)}\rangle$ is unitary. We then
have two constraint operators, $\hat{C}$ and $\hat{T}$, but since $\hat{T}$ is
a Casimir operator and commutes with all constraints it is easier to solve
\cite{Casimir}. (The Casimir property also implies that $\hat{C}$ and
$\hat{T}$ form a pair of first-class constraints.) There is no gauge flow
associated with effective constraints of $\hat{T}$, but only relations between
moments. For instance, for second-order moments we have a general relationship
$\langle\widehat{\sin(\delta P)} \rangle \Delta(\sin(\delta P)\cdots)+
\langle\widehat{\cos(\delta P)} \rangle \Delta(\cos(\delta P)\cdots)=0$ which
allows us to eliminate all moments involving $\cos(\delta P)$. See
App.~\ref{a:Casimir} for more details.

We will be
interested in second-order moments, for which we include
\begin{eqnarray} \label{Ceff}
\langle\hat{C}\rangle&=& \delta^2\langle\hat{p}_{\lambda}\rangle^2-
\langle\hat{Q}\rangle^2 \langle\widehat{\sin(\delta P)}\rangle^2+ \delta^2
\Delta(p_{\lambda}^2)- \langle\widehat{\sin(\delta P)}\rangle^2 \Delta(Q^2)-
\langle\hat{Q}\rangle^2 \Delta(\sin(\delta P)^2) \nonumber\\
&&-
4\langle\hat{Q}\rangle\langle\widehat{\sin(\delta P)}\rangle \Delta(Q\sin(\delta
P))
\end{eqnarray}
and constraints $C_{\rm pol}$ with linear polynomials. (We ignore re-ordering
terms which would have explicit factors of $\hbar$.) We will not need
$C_{\lambda}$ here, but do use
\begin{eqnarray}
 C_{p_{\lambda}} &=& 2\delta^2\langle\hat{p}_{\lambda}\rangle
 \Delta(p_{\lambda}^2)- 2\langle\hat{Q}\rangle
 \langle\widehat{\sin(\delta P)}\rangle \Delta(p_{\lambda}Q)-
 2\langle\hat{Q}\rangle^2 \langle\widehat{\sin(\delta P)}\rangle
 \Delta(p_{\lambda}\sin(\delta P))\label{Cpl}\\
 C_Q &=& 2\delta^2\langle\hat{p}_{\lambda}\rangle \Delta(p_{\lambda}Q)-
 2\langle\hat{Q}\rangle \langle\widehat{\sin(\delta P)}\rangle^2 \Delta(Q^2)
 \nonumber\\ 
 &&-
 2\langle\hat{Q}\rangle^2 \langle\widehat{\sin(\delta P)}\rangle
 \left(\Delta(Q\sin(\delta P))+
   \frac{1}{2}i\hbar\delta\langle\widehat{\cos(\delta P)}\rangle\right) \label{CQ}\\ 
 C_{\sin(\delta P)} &=& 2\delta^2\langle\hat{p}_{\lambda}\rangle
 \Delta(p_{\lambda}\sin(\delta P))- 2\langle\hat{Q}\rangle \langle\sin(\delta
 \hat{P})\rangle^2 \left(\Delta(Q\sin(\delta P))-
   \frac{1}{2}i\hbar\delta\langle\widehat{\cos(\delta P)}\rangle\right)\nonumber\\
&&-
 2\langle\hat{Q}\rangle^2 \langle\widehat{\sin(\delta P)}\rangle \Delta(\sin(\delta
 P)^2)\,. \label{CP}
\end{eqnarray}
If we solve $C_Q=0$ and $C_{\sin(\delta P)}=0$ for $\Delta(p_{\lambda}Q)$ and
$\Delta(p_{\lambda}\sin(\delta P))$ and insert the results in the solution for
$\Delta(p_{\lambda}^2)$ obtained from $C_{p_{\lambda}}=0$, we find
\begin{equation} \label{Deltapl}
 \Delta(p_{\lambda}^2) = \langle\hat{p}_{\lambda}\rangle^2
 \frac{\Delta(Q^2)}{\langle\hat{Q}\rangle^2}+
 \frac{2\langle\hat{p}_{\lambda}\rangle}{\delta} \Delta(Q\sin(\delta P))+
 \frac{\langle\hat{Q}\rangle^2}{\delta^2} \Delta(\sin(\delta P)^2)
\end{equation}
using that (\ref{Ceff}) implies
$\langle\hat{Q}\rangle^2\langle\widehat{\sin(\delta P)}\rangle^2=\delta^2
\langle\hat{p}_{\lambda}\rangle^2+O(\Delta(\cdot))$ and suppressing quadratic
terms in second-order moments along with higher-order ones. (Owing to
imaginary contributions in the effective constraints $C_Q$ and $C_{\sin(\delta
  P)}$, the moments $\Delta(p_{\lambda}Q)$ and $\Delta(p_{\lambda}\sin(\delta
P))$ are complex. We discuss this feature in the appendix.)

\paragraph{Gaussian in the volume.}

For an uncorrelated Gaussian
\begin{equation}
 \psi(Q)= \frac{1}{\sqrt[4]{2\pi}\sqrt{\sigma}}
 \exp\left(-\frac{(Q-\bar{Q})^2}{4\sigma^2}\right) \exp(i\bar{P}Q/\hbar)\,,
\end{equation}
we have $\Delta(Q\sin(\delta P))=0$ and
\begin{equation}
  \delta^{-2}\Delta(\sin(\delta P)^2)= \langle\widehat{\cos(\delta P)}\rangle^2
  \Delta(P^2)=
  \left(1-\frac{\delta^2
      \langle\hat{p}_{\lambda}\rangle^2}{\langle\hat{Q}\rangle^2}
 +O(\Delta(\cdot))\right) 
  \frac{\hbar^2}{4\Delta(Q^2)}\,.
\end{equation}
In this case, using (\ref{Deltapl}),
\begin{equation}
 \Delta p_{\lambda} = \sqrt{\frac{\hbar^2}{4}
   \frac{\langle\hat{Q}\rangle^2}{\Delta(Q^2)}+
   \langle\hat{p}_{\lambda}\rangle^2
   \left(\frac{\Delta(Q^2)}{\langle\hat{Q}\rangle^2}-
     \frac{\delta^2\hbar^2}{4\Delta(Q^2)}\right)}
\end{equation}
or (for $x=-1/2$)
\begin{equation} \label{DeltapphiV}
 \Delta p_{\phi} = \sqrt{3\pi G\hbar^2
   \frac{\langle\hat{V}\rangle^2}{(\Delta V)^2}+
   \langle\hat{p}_{\phi}\rangle^2
   \left(\frac{(\Delta V)^2}{\langle\hat{V}\rangle^2}-
     \frac{4\pi^2\delta^2\ell_{\rm P}^4}{(\Delta V)^2}\right)}\,.
\end{equation}

\begin{figure}
\begin{center}
\includegraphics[width=10cm]{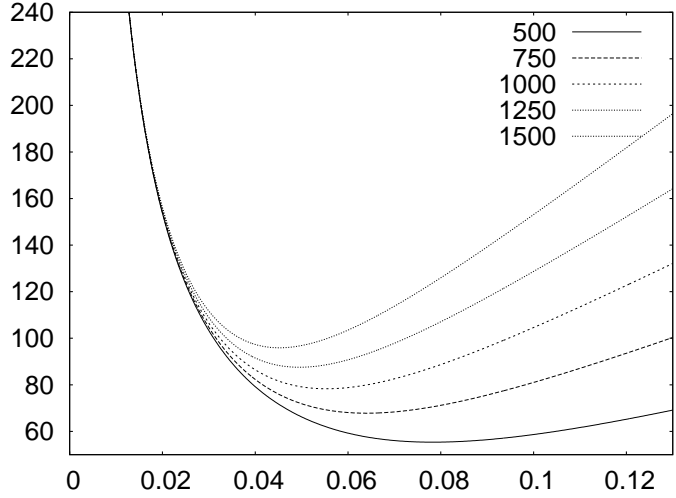}
\caption{The $p_{\phi}$-fluctuation (\ref{DeltapphiV}) as a function of
  $\Delta V/\langle\hat{V}\rangle$, for different values of $p_{\phi}$. Other
  parameters are as in Fig.~\ref{Fig:GaussianV}.  This plot is to be compared
  with Fig.~11 of \cite{NumBounce}. \label{Fig:DeltaVV}}
\end{center}
\end{figure}

The last contribution can be written as $-3\pi G\hbar^2 (\rho_{\rm
  eff}/\rho_{\rm QG}) \langle\hat{V}\rangle^2/(\Delta V)^2$ and is therefore
small compared to the first term for an initial state. Ignoring the last
contribution, $\Delta p_{\phi}$ is plotted as a function of $\Delta
V/\langle\hat{V}\rangle$ in Fig.~\ref{Fig:DeltaVV}.  For small $\Delta
V/\langle\hat{V}\rangle$, the first term is dominant and $\Delta p_{\phi}$ is
independent of $\langle\hat{p}_{\phi}\rangle$. We can also see that
\begin{equation}
 \frac{\Delta V}{\langle\hat{V}\rangle} \Delta p_{\phi}= \sqrt{3\pi G\hbar^2+
   \langle\hat{p}_{\phi}\rangle^2\left(\frac{\Delta
       V}{\langle\hat{V}\rangle}\right)^4- 4\pi^2\delta^2\ell_{\rm P}^4
   \frac{\langle\hat{p}_{\phi}\rangle^2}{\langle\hat{V}\rangle^2}}
\end{equation}
is bounded from below by $\sqrt{3\pi G}\hbar$, noting that the last term is
small when the volume expectation value is larger than Planckian. The minimum
value $(\Delta p_{\phi})_{\rm min}^2= 2\sqrt{3\pi G}\hbar p_{\phi}$ is
obtained for $(\Delta V/\langle\hat{V}\rangle)^2= \sqrt{3\pi
  G}\hbar/p_{\phi}$. Some of these limiting cases have been mentioned in
\cite{NumBounce}.

\begin{figure}
\begin{center}
\includegraphics[width=10cm]{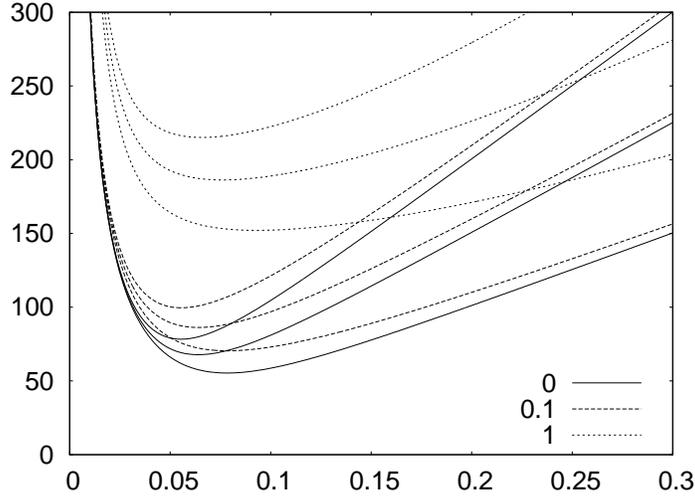}
\caption{The $p_{\phi}$-fluctuation (\ref{DeltapphiVCorr}) as a function of
  $\Delta V/\langle\hat{V}\rangle$, for different values of $p_{\phi}$ and
  $\kappa$ (as shown in units of $\hbar$). Other parameters are as in
  Fig.~\ref{Fig:GaussianV}.  \label{Fig:DeltaVVC}}
\end{center}
\end{figure}

For a correlated Gaussian, we obtain
\begin{equation} \label{DeltapphiVCorr}
 \Delta p_{\phi} = \sqrt{3\pi G\hbar^2 (1+4\kappa^2/\hbar^2)
   \frac{\langle\hat{V}\rangle^2}{\Delta(V^2)} \left(1-\frac{\rho_{\rm
         free}}{\rho_{\rm QG}}\right)+
   \langle\hat{p}_{\phi}\rangle^2
   \left(\frac{\Delta(V^2)}{\langle\hat{V}\rangle^2}
     +2\frac{\Delta(VP)}{\langle\hat{V}\rangle\langle\hat{P}\rangle}\right)}\,. 
\end{equation}
At small curvature (appropriate for an initial state at large volume), we can
write the last contribution as $2\langle\hat{p}_{\phi}\rangle^2
\Delta(VP)/\langle\hat{V}\rangle\langle\hat{P}\rangle)=4\sqrt{3\pi G\hbar^2}
\langle\hat{p}_{\phi}\rangle \kappa/\hbar$. In this form, the behavior of $\Delta
p_{\phi}$ is shown in Fig.~\ref{Fig:DeltaVVC}. Effects of $\kappa\not=0$ are
largest around the minimum of $\Delta V/\langle\hat{V}\rangle$ as a function
of $\Delta p_{\phi}$.

\paragraph{Gaussian in the scalar.}

A state in the volume representation which is Gaussian for the field value is
of the form (\ref{phiGaussian}). In addition to expectation values and moments
computed before, we have
\begin{equation}
  \langle\widehat{\sin(\delta P)}\rangle = \frac{\delta
    \bar{p}_{\lambda}}{\bar{Q}} e^{1/2\sigma^2}
\end{equation}
and 
\begin{eqnarray}
 \Delta(Q^2) &=& \bar{Q}^2(e^{2/\sigma^2}-e^{1/\sigma^2})=
 \langle\hat{Q}\rangle^2(e^{1/\sigma^2}-1)\\
\Delta(Q\sin(\delta P)) &=& \delta \bar{p}_{\lambda}
(1-e^{1/\sigma^2})\\
\Delta(\sin(\delta P)^2) &=& 
3\frac{\delta^2\hbar^2}{\bar{Q}^2} \left(\frac{\sigma^2}{4} +1\right)
e^{2/\sigma^2} + \frac{\delta^2\bar{p}_{\lambda}^2}{\bar{Q}^2} \left(4(1+4/\sigma^4)
  e^{2/\sigma^2}-1\right)\,.
\end{eqnarray}

Our moment relations derived from effective constraints then imply
\begin{equation}
  \Delta p_{\phi} = \sqrt{36\pi G\hbar^2 \left(\frac{\sigma^2}{4}+1\right)
    e^{3/\sigma^2} +\langle\hat{p}_{\phi}\rangle^2
    \left(4\left(1+\frac{4}{\sigma^2}\right) e^{3/\sigma^2}-
      2e^{1/\sigma^2}+1\right)}\,.
\end{equation}
For small $\Delta V/\langle\hat{V}\rangle$ (large $\sigma$), the most relevant
term is $\Delta (p_{\phi})^2\sim 9\pi G\hbar^2 \sigma^2\sim 9\pi
G\hbar^2 (\langle\hat{V}\rangle/\Delta V)^2$, which is independent of
$\langle\hat{p}_{\phi}\rangle$. Moreover, $\Delta p_{\phi}$ is inversely
proportional to $\Delta V/\langle\hat{V}\rangle$. 

\begin{figure}
\begin{center}
\includegraphics[width=10cm]{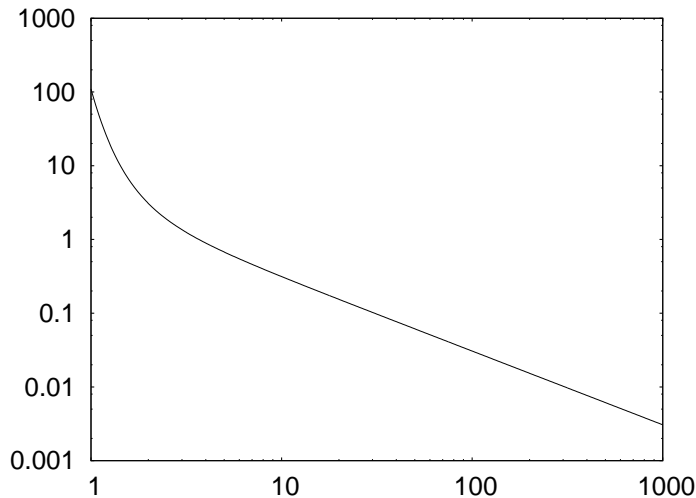}
\caption{The relative volume fluctuation (\ref{DeltaVphi}) as a function of
  $\Delta p_{\phi}$, for arbitrary values of $p_{\phi}$. Other
  parameters are as in Fig.~\ref{Fig:GaussianV}.  This plot is to be compared
  with Fig.~16 of \cite{NumBounce}. \label{Fig:DeltaVphi}}
\end{center}
\end{figure}

As noted before, we must be careful with the expansions used here for
integrations when volume fluctuations are not sufficiently small. Fortunately,
the relation between $\Delta V$ and $\Delta p_{\phi}$ can be obtained more
easily from a combination of Eqs.~(\ref{Deltalambda}) and (\ref{DeltaQ}): With
$\sigma=2\Delta p_{\lambda}/\hbar= (3\pi G\hbar^2)^{-1/2} \Delta p_{\phi}$, we
obtain
\begin{equation} \label{DeltaVphi}
\frac{\Delta V}{\langle\hat{V}\rangle} = \sqrt{\exp\left(\frac{3\pi
      G\hbar^2}{(\Delta p_{\phi})^2}\right) -1}
\end{equation}
which equals $\exp(\frac{3}{2}\pi G\hbar^2/(\Delta p_{\phi})^2)$ for $\Delta
p_{\phi}\ll 3\pi G\hbar^2$, and $\sqrt{3\pi G\hbar^2}/\Delta p_{\phi}$ for
$\Delta p_{\phi}\gg 3\pi G\hbar^2$. The full function is shown in
Fig.~\ref{Fig:DeltaVphi}. Notice that $\Delta p_{\phi} \Delta
V/\langle\hat{V}\rangle$ is a constant (equal to $\sqrt{3\pi G\hbar^2}$) for
small relative volume fluctuations, so that setting $\Delta
Q/\langle\hat{Q}\rangle\approx \Delta\lambda$ is consistent with the
saturation of the $(\lambda,p_{\lambda})$-uncertainty relation for the
Gaussian (\ref{phiGaussian}), as remarked in \cite{NumBounce}. For large
values, $\Delta Q/\langle\hat{Q}\rangle$ is no longer proportional to
$\Delta\lambda$ as the classical solutions, identifying $\lambda$ with
$\log(Q/\bar{Q})+\bar{\lambda}$, would suggest. The minimal-uncertainty
Gaussian in the scalar therefore may lead to larger values of $\Delta
V/\langle\hat{V}\rangle$ than expected from an identification of this relative
fluctuation with $\Delta\lambda$.

For a squeezed state, we have the previous expression (\ref{DeltaQ}) for
$\Delta Q$ in terms of $\sigma$, but now $\Delta
p_{\lambda}=\frac{1}{2}\hbar\sigma\sqrt{1+4\kappa^2/\hbar^2}$. Thus,
\begin{equation}
\frac{\Delta V}{\langle\hat{V}\rangle} = \sqrt{\exp\left(\frac{3\pi
      G\hbar^2}{(\Delta p_{\phi})^2}(1+4\kappa^2/\hbar^2)\right) -1}\,.
\end{equation}
Different values of $\kappa\not=0$ simply shift the curve in
Fig.~\ref{Fig:DeltaVphi} up.

\subsubsection{Density operator}

The fluctuation energy is relevant for the maximum density reached along
solutions to the Hamiltonian constraint of the harmonic model. According to
(\ref{epsilonQ}), departures from $\rho_{\rm QG}$ are largest for small volume
fluctuations. However, a second source of volume fluctuations matters for a
comparison of evolved densities with the value $\rho_{\rm QG}$: Depending on
what expression is used to compute density expectation values, additional
moments may appear as in (\ref{Dens}).

As shown in \cite{DensityOp}, the density operator $\hat{\rho}=\frac{1}{2}
\hat{p}_{\phi}^2\hat{V}^{-2}$ has a continuous spectrum bounded by $\rho_{\rm
  QG}$ in models of loop quantum cosmology, irrespective of the precise factor
ordering of the constraint. Although the expectation value
$\langle\hat{\rho}\rangle$ in a specific evolved state may not necessarily
reach this bound (or may even be larger if there are discrete eigenvalues of
$\hat{\rho}$ above $\rho_{\rm QG}$), the value of $\rho_{\rm QG}$ sets a
distinguished scale in this problem.

Effective equations, such as (\ref{EffFried}), contain density
parameters. However, as seen from the detailed derivation in
Section~\ref{s:Eff}, these effective densities are defined as $\rho_{\rm
  eff}=\frac{1}{2} \langle\hat{p}_{\phi}\rangle^2/\langle\hat{V}\rangle^2$,
not as $\langle\hat{\rho}\rangle$.  We can then write (\ref{Dens}) as an
expression that relates the effective density to the density expectation
value:
\begin{equation}
  \rho_{\rm eff} = \frac{\langle\hat{\rho}\rangle}{
1+\frac{\Delta(p_{\phi}^2)}{\langle\hat{p}_{\phi}\rangle^2}+
   3\frac{\Delta(V^2)}{\langle\hat{V}\rangle^2}-
   4\frac{\Delta(Vp_{\phi})}{\langle\hat{p}_{\phi}\rangle\langle\hat{V}\rangle}
+\cdots}\,.
\end{equation}
We ignore higher-order moments for semiclassical states, but they may
contribute for larger volume fluctuations.

The remaining moments are all related to the volume fluctuation, as shown in
the preceding subsection: For an uncorrelated $V$-Gaussian,
\begin{eqnarray}
 \frac{(\Delta p_{\phi})^2}{\langle\hat{p}_{\phi}\rangle^2} &=&
 \frac{(\Delta V)^2}{\langle\hat{V}\rangle^2}+ 
 \left(\frac{3\pi G\hbar^2}{\langle\hat{p}_{\phi}\rangle^2}-
\frac{4\pi^2\delta^2\ell_{\rm P}^4}{\langle\hat{V}\rangle^2} \right)
 \frac{\langle\hat{V}\rangle^2}{(\Delta V)^2}\\
 \frac{\Delta(Vp_{\phi})}{\langle\hat{V}\rangle \langle\hat{p}_{\phi}\rangle} 
&=& \frac{(\Delta V)^2}{\langle\hat{V}\rangle^2}\,.
\end{eqnarray}
Thus,
\begin{equation}
  \rho_{\rm eff} = \frac{\langle\hat{\rho}\rangle}{
1-4\pi^2\frac{\delta^2\ell_{\rm P}^4}{(\Delta V)^2}+ 
 \frac{3\pi G\hbar^2}{\langle\hat{p}_{\phi}\rangle^2}
\frac{\langle\hat{V}\rangle^2}{(\Delta V)^2}
+\cdots}
=\frac{\langle\hat{\rho}\rangle}{1-\epsilon\left(\frac{\rho_{\rm
        QG}}{\rho_{\rm eff}}-1\right)}
\end{equation}
for such a state. This equation can be solved for $\rho_{\rm eff}$ and implies
$\rho_{\rm eff}(1+\epsilon)=\langle\hat{\rho}\rangle+\epsilon \rho_{\rm QG}$.
If $\langle\hat{\rho}\rangle^{\rm max}=\rho_{\rm QG}$, $\rho_{\rm eff}^{\max}=
\rho_{\rm QG}$. If $\langle\hat{\rho}\rangle^{\rm max}=\rho_{\rm
  QG}(1+\epsilon)$, $\rho_{\rm eff}^{\max}= \rho_{\rm
  QG}(1+2\epsilon)/(1+\epsilon)\sim\rho_{\rm QG}(1+\epsilon)$.  With
correlations, we obtain
\begin{equation}
 \rho_{\rm eff} = \frac{\langle\hat{\rho}\rangle}{1-\epsilon(1+4\kappa^2/\hbar^2)
   (\rho_{\rm QG}/\rho_{\rm eff}-1)-
   2\Delta(VP)/\langle\hat{V}\rangle\langle\hat{P}\rangle}
\end{equation}
If we assume $\rho^{\rm max}=\rho_{\rm QG}$, $\rho_{\rm eff}^{\rm
  max}=\rho_{\rm QG}/(1-2(1+\epsilon(1+4\kappa^2/\hbar^2))
\Delta(VP)/\langle\hat{V}\rangle\langle\hat{P}\rangle)$ is not restricted to
be less than $\rho_{\rm QG}$.

For a Gaussian in the scalar, the moments already computed give us
\begin{eqnarray}
  \frac{(\Delta V)^2}{\langle\hat{V}\rangle^2} &=& e^{1/\sigma^2}-1\\
 \frac{\Delta(p_{\phi}V)}{\langle\hat{p}_{\phi}\rangle \langle\hat{V}\rangle}
 &=& e^{3/2\sigma^2} (e^{1/\sigma^2}-1)\\
 \frac{(\Delta p_{\phi})^2}{\langle\hat{p}_{\phi}\rangle^2} &=&
 4\left(1+\frac{4}{\sigma^2}\right) e^{3/\sigma^2}- 2e^{1/\sigma^2}+1+
 \frac{36\pi G\hbar^2}{\langle\hat{p}_{\phi}\rangle^2}
 \left(\frac{\sigma^2}{4}+1\right) e^{3/\sigma^2}\,.
\end{eqnarray}
The first relation can be inverted for 
\begin{equation}
 \sigma=\frac{1}{\sqrt{\log(1+(\Delta V/\langle\hat{V}\rangle)^2)}}\,.
\end{equation}
The final expression for $\rho_{\rm eff}$ in terms of
$\langle\hat{\rho}\rangle$ is lengthy.
We do not write it here because we have to be careful with our approximate
integrals for large relative moments.

\begin{figure}
\begin{center}
\includegraphics[width=9.7cm]{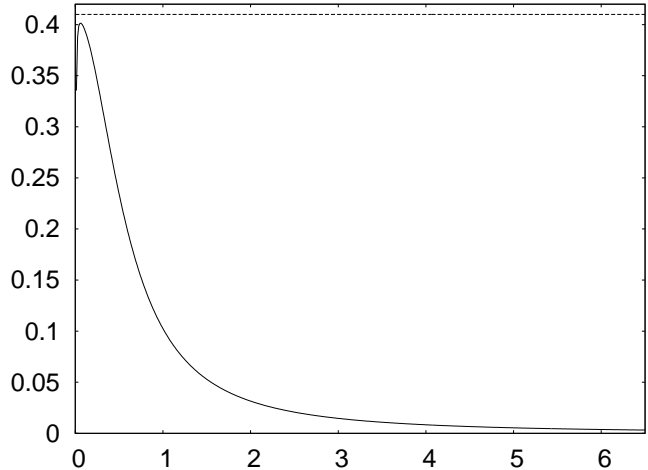}
\caption{The effective density (\ref{rhophi2}) (for $\kappa=0$) as a function of
  $\Delta V/\langle\hat{V}\rangle$, assuming $\langle\hat{\rho}\rangle^{\rm
    max}=\rho_{\rm QG}$. This plot qualitatively agrees with Fig.~20 of
  \cite{NumBounce}. The quantitative difference can be explained well by the
  fact that we take volume fluctuations and the effective density at the same
  time, while \cite{NumBounce} refers to initial volume
  fluctuations. Moreover, $\langle\hat{\rho}\rangle^{\rm max}$ may differ from
  $\rho_{\rm QG}$ in a way depending on $\Delta V/\langle\hat{V}\rangle$. The
  features visible in this plot are independent of
  $\langle\hat{p}_{\phi}\rangle$; see Fig.~\ref{Fig:GaussianphiC} for the
  $\langle\hat{p}_{\phi}\rangle$-dependence with small $\Delta
  V/\langle\hat{V}\rangle$. \label{Fig:Gaussianphi}}
\end{center}
\end{figure}

\begin{figure}
\begin{center}
\includegraphics[width=9.7cm]{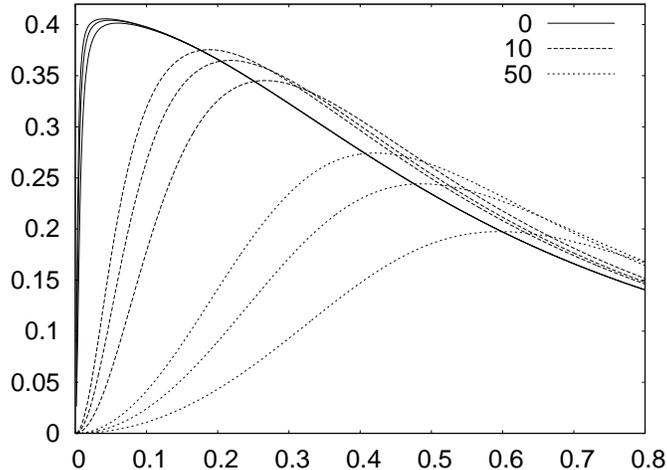}
\caption{The effective density (\ref{rhophi2}) as a function of $\Delta
  V/\langle\hat{V}\rangle$, assuming $\langle\hat{\rho}\rangle^{\rm
    max}=\rho_{\rm QG}$. The maximum density obtained decreases for growing
  covariance. For each of the three displayed values of $\kappa$ (in units
  with $\hbar=1$), three curves for $\langle\hat{p}_{\phi}\rangle\in
  \{500,750,1000\}$ are shown. (Rather extreme values have been chosen for
  $\kappa/\hbar$ in order to bring out the dependence on this value more
  clearly. For $\kappa/\hbar\sim O(1)$, the effective density does not vary
  much with $\kappa/\hbar$.) \label{Fig:GaussianphiC}}
\end{center}
\end{figure}

We can compute $\Delta(p_{\lambda}Q)$ more easily by using the
$\hat{p}_{\lambda}$-operator, which gives us
$\langle\hat{Q}\hat{p}_{\lambda}\rangle=
\langle\hat{Q}\rangle(\frac{1}{2}i\hbar+\kappa+\bar{p}_{\lambda})$, and then
$\Delta(p_{\lambda}Q)= {\rm Re}\langle\hat{Q}\hat{p}_{\lambda}\rangle-
\langle\hat{Q}\rangle\langle\hat{p}_{\lambda}\rangle= \kappa$. With our previous
result for $\Delta p_{\lambda}$, we have
\begin{equation} \label{rhophi2}
 \rho_{\rm eff} = \frac{\langle\hat{\rho}\rangle}{1+\frac{3\pi
     G\hbar^2}{\langle\hat{p}_{\phi}\rangle^2}
   \left(1+4\frac{\kappa^2}{\hbar^2}\right) \frac{1}{\log(1+(\Delta
     V)^2/\langle\hat{V}\rangle ^2)} + 3\frac{(\Delta
     V)^2}{\langle\hat{V}\rangle ^2} - 4\frac{\sqrt{3\pi
       G\hbar^2}}{\langle\hat{p}_{\phi}\rangle} \frac{\kappa}{\hbar}}\,.
\end{equation}
Assuming a maximum $\langle\hat{\rho}\rangle$ given by $\rho_{\rm QG}$,
(\ref{rhophi2}) as a function of $\Delta V/\langle\hat{V}\rangle$ is shown in
Fig.~\ref{Fig:Gaussianphi} for $\kappa=0$ and in Fig.~\ref{Fig:GaussianphiC}
for different values of $\kappa$. For large $\kappa/\hbar$, the maximum
effective density depends on $\kappa$ and $\langle\hat{p}_{\phi}\rangle$
mainly through the combination $\kappa/\langle\hat{p}_{\phi}\rangle$.

\section{Conclusions}

We have provided several applications of effective equations and constraints
in quantum cosmology, focusing on background properties that are important for
instabilities of initial-value problems in Euclidean regimes and on relations
between moments that affect density bounds. The latter results provide
fruitful comparisons with recent numerical evolutions of wave functions
\cite{NumBounce}, showing the agreement for effective equations that include
moment terms. While \cite{NumBounce} compared numerical results only with
zeroth-order effective equations without moment terms, finding several
disagreements, our application of existing effective methods show that
moment-dependent corrections capture the evolution of wave functions to an
excellent degree. In particular, the relations between moments derived here
show that properties of physical states can be computed reliably by effective
methods without having to enter technical intricacies usually associated with
physical Hilbert spaces.

This conclusion has several implications for the interpretation of quantum
cosmology at high density. For some time, a universal effective equation of
the form (\ref{ModFried}) has been claimed to capture the evolution of a
significant class of states and matter models in loop quantum cosmology. The
combined results of effective equations, accumulated over several years, and
recent numerical simulations shows that the equation is unreliable when
moments become large and states are no longer sharply peaked. For small volume
fluctuations in a Gaussian (and correspondingly large momentum fluctuations)
the correction term has been derived in \cite{BouncePert} and is one of the
terms used here to show agreement with numerical simulations. For large volume
fluctuations, deviations from the maximum density predicted by
(\ref{ModFried}) can be explained by effective equations as well, but since
their validity to second order in moments is no longer clear, the effect was
brought out clearly only by the numerical simulations of
\cite{NumBounce}. Nevertheless, even for such large volume fluctuations there
is, rather surprisingly, good agreement.

Effective equations and their implications are therefore reliable. One of
these implications at a general level is the fact that evolution and physical
properties at high density depend sensitively on the initial quantum state
used, in a way that is difficult to control in quantum cosmology. The examples
of states used in \cite{NumBounce} and here make this clear, even though these
states are rather limited as well. Both classes of states are Gaussian, one in
the geometrical variables $(Q,P)$ and one in the scalar pair
$(\phi,p_{\phi})$, and therefore much restricted even compared with a general
class of semiclassical states. Nevertheless, even these restricted classes of
states show marked differences for large volume fluctuations, with only small
effects on the maximum density for $(Q,P)$-Gaussians and a strong suppression
of the maximum density for $(\phi,p_{\phi})$-Gaussians. One might argue that
large relative volume fluctuations should not occur for a good universe model
if they refer to a semiclassical initial state at low curvature (as in
\cite{NumBounce}). However, in the free, massless model there is not much
quantum back-reaction, and by assuming a sharply-peaked initial state one
implicitly assumes a sharply-peaked state at high density. In more-general
models, relative volume fluctuations can easily grow, so that states with
large relative fluctuations should indeed be considered for a generic quantum
phase. Then, there is a strong sensitivity on the precise state used.

This result highlights the fact that loop quantum cosmology at large density
cannot be used for reliable predictions, unless one obtains a much better
understanding of relevant quantum states and possible conditions one can
impose on them independently of the constraints. (The authors of
\cite{NumBounce} do not arrive at the same conclusion. Moreover, by referring
to the different state choices as different ``methods'' of implementing
initial states numerically, they hide the fact that different initial states
imply different physics, and ultimately different predictions.)

We finally highlight a small but important result which has not played a large
role in our considerations so far: As shown by Eq.~(\ref{EpsilonCancel}), for
some states the fluctuation energy may cancel completely the modification of
the classical Friedmann equation that results from holonomy effects. Although
this is only one example, it shows that moment terms in effective equations
can easily be of the same order as holonomy corrections. Fluctuation energies
considered here or, more importantly, quantum back-reaction terms in
non-harmonic cosmological models with a massive or or self-interacting scalar,
are therefore crucial for the task of evaluating the reliability of scenarios
based on holonomy effects. (This result agrees with the expectation that
higher-curvature corrections generically appear in effective actions of
gravity: Higher time derivatives in the corresponding effective equations
result from quantum back-reaction of moments \cite{HigherTime}. Moment terms
should then be of the same order $O(\ell_{\rm P}^2{\cal H}^2)$ as
higher-curvature corrections, which is the same order as holonomy
modifications obtained by expanding the sine function in (\ref{Cmod}).)
Unfortunately, not much is known about the generic behavior of quantum
back-reaction in (loop) quantum cosmology. The good agreement of effective and
numerical calculations confirmed in this article will hopefully lead to a
more-complete understanding.

\section*{Acknowledgements}

The author thanks Jaume Garriga for discussions, pointing out the important
question of quantitative aspects of instabilities in Euclidean regimes. This
work was supported in part by NSF grant PHY-1307408.

\begin{appendix}
\section{Kinematical and physical moments}

If we solve (\ref{CQ}) and (\ref{CP}) for $\Delta(p_{\lambda}Q)$ and
$\Delta(p_{\lambda}\sin(\delta P))$, respectively, we obtain complex values
\begin{eqnarray} \label{ComplexMoments}
 \delta\Delta(p_{\lambda}Q) &=& \langle\widehat{\sin(\delta P)}\rangle \Delta(Q^2)+
 \langle\hat{Q}\rangle \Delta(Q\sin(\delta P))+ \frac{1}{2} i\hbar\delta
 \langle\hat{Q}\rangle \langle\widehat{\cos(\delta P)}\rangle\\
\delta\Delta(p_{\lambda}\sin(\delta P)) &=& \langle\sin(\delta
\hat{P})\rangle\Delta(Q\sin(\delta P))+
\langle\hat{Q}\rangle\Delta(\sin^2(\delta P))- \frac{1}{2} i\hbar\delta
 \langle\hat{Q}\rangle \langle\widehat{\cos(\delta P)}\rangle
 \langle\widehat{\sin(\delta P)}\rangle \,. \nonumber
\end{eqnarray}
(As before, we have used $\delta\langle\hat{p}_{\lambda}\rangle=
\langle\hat{Q}\rangle \langle\widehat{\sin(\delta P)}\rangle$ in coefficients
of the moments, which is valid to the order considered here since squared
relative moments are then small compared to the relative moments.) The
imaginary terms cancel in the combination (\ref{Deltapl}) which was of
interest earlier on in this article, but it is still important and instructive
to comment on the complex-valuedness of some moments. This feature is related
to the complex nature of time expectation values in deparameterized quantum
systems, noticed in \cite{EffTime,EffTimeLong} and evaluated in a cosmological
model in \cite{EffTimeCosmo}. We will first dicuss complex moments and their
physical role in detail, and then apply the notion of complex time to the
models studied here.

\subsection{Physical states in effective constrained systems}

The moments appearing in (\ref{ComplexMoments}) involve all phase-space
variables (except $\lambda$) of the original system with degrees of freedom
$(Q,P)$ and $(\lambda,p_{\lambda})$. These are independent variables only
before constraints are solved. Their quantum moments can be computed only in a
kinematical Hilbert space of states that do not solve the constraints. Since
moments refer to symmetric orderings of the operators involved, they must take
real values when computed in the kinematical Hilbert space.

In (\ref{ComplexMoments}), we have solved some of the effective constraints
and therefore left the kinematical arena. When all constraints are imposed at
the quantum level, we have moments of physical states in which some of the
kinematical degrees of freedom have been removed. We can no longer consider
$(Q,P)$ and $(\lambda,p_{\lambda})$ as independent, and their expectation
values and moments are related according to the effective constraints. This
observation already helps to explain why imaginary contributions may occur in
solutions for some of these moments: While $\hat{p}_{\lambda}$ and $\hat{Q}$
quantize independent degrees of freedom and commute with each other as
operators on the kinematical Hilbert space, they are related and may be
non-commuting on the physical Hilbert space. (Indeed, in the present example,
$\hat{p}_{\lambda}\psi=\delta^{-1} \widehat{Q\sin(\delta P)}\psi$ on physical
states.)

\subsubsection{Physical moments}

Transitioning from the kinematical to a physical Hilbert space is usually a
complicated procedure in canonical quantum gravity and cosmology, and indeed
presents one of the crucial problems to be faced. One can easily derive
physical Hilbert spaces in deparameterizable systems, forming by far the
largest class of models studied in this context. However, the special nature
of deparameterizable systems, requiring the existence of a phase-space
variable without turning points where the momentum becomes zero, makes it
doubtful that results relying on deparameterizability are robust. One of the
advantages of effective constraints is that they can be applied consistently
to systems with local internal times at the quantum level.

Effective methods allow one to compute properties of quantum states without
having to work with wave functions and explicit integral representations of
inner products in Hilbert spaces. Instead, states are expressed in terms of
the expectation-value functional, or the set of expectation values and moments
of basic operators. As successfully used in algebraic quantum field theory
\cite{LocalQuant}, for instance, a state is a positive linear functional of
norm one on the $*$-algebra generated by basic operators. If one uses a
Hilbert space, the $*$-relation is turned into adjointness relations of
operators. The positive linear functional applied to an element $A$ of the
$*$-algebra is then the expectation value of a given wave function assigned to
$A$, computed using the inner product of the Hilbert space.  It turns out that
many questions of interest do not require much of the structure of a Hilbert
space but rather refer to general properties of states as positive linear
functionals. These are exactly the properties that can be computed using
canonical effective methods. (Properties that are difficult to express in this
way refer to probabilities of individual measurements. Such questions rarely
play a role in quantum cosmology.)

Physical normalization, in the absence of an explicit Hilbert space, is
imposed by requiring all observables --- expectation values and moments --- to
respect the $*$-relations of the $*$-algebra. For the usual canonical basic
operators generating a $*$-algebra, this means that all expectation values and
moments must be real. This simple condition is sufficient to replace
complicated constructions of integral representations of physical inner
products. (Positivity of the linear functional of a state is reflected in
inequalities to be satisfied by physical moments, most importantly uncertainty
relations.) The general notion of states as positive linear functionals also
makes it possible to formulate a unified treatment of kinematical and physical
moments as used here. Both types of moments are defined on the same
$*$-algebra, but physical moments are restricted by (i) reality conditions
that follow from the $*$-relation and (ii) by effective constraints derived
from constraint operators. Even when it is not possible to find a physical
Hilbert space as a subset of the kinematical Hilbert space, kinematical and
physical moments can be treated on the same footing. The remainder of this
appendix provides new examples for this scheme.

\subsubsection{Casimir constraints}
\label{a:Casimir}

Casimir constraints are special versions of first-class constraints which do
not generate a gauge flow (which is possible for first-class constraints on
non-symplectic spaces). They lead to restrictions of the moments, but no flow
need be factored out. Unrestricted moments are therefore automatically Dirac
observables. 

The condition (\ref{T}) provides an example for Casimir constraints. It
implies an effective constraint
\[
 \langle\hat{T}\rangle= \langle\widehat{\sin(\delta P)}\rangle^2+
 \langle\widehat{\cos(\delta P)}\rangle^2 +\Delta(\sin^2(\delta P))+
 \Delta(\cos^2(\delta P))-1=0
\]
as well as higher-order constraints $T_{\rm pol}$. We will make use only of 
\begin{eqnarray*}
 T_{\sin(\delta P)} &=& 2\langle\widehat{\sin(\delta P)}\rangle
 \Delta(\sin^2(\delta P)+ 2\langle\widehat{\cos(\delta P)}\rangle
 \Delta(\sin(\delta P)\cos(\delta P))=0\\
 T_{\cos(\delta P)} &=& 2\langle\widehat{\sin(\delta P)}\rangle
 \Delta(\sin(\delta P)\cos(\delta P))+ 2\langle\widehat{\cos(\delta P)}\rangle
 \Delta(\cos^2(\delta P))\,.
\end{eqnarray*}
We can eliminate the two moments involving $\cos(\delta P)$:
\begin{eqnarray*}
 \Delta(\sin(\delta P)\cos(\delta P)) &=& \frac{\langle\widehat{\sin(\delta
     P)}\rangle}{\langle\widehat{\cos(\delta P)}\rangle} \Delta(\sin^2(\delta
 P))\\
 \Delta(\cos^2(\delta P)) &=& \frac{\langle\widehat{\sin(\delta
     P)}\rangle^2}{\langle\widehat{\cos(\delta P)}\rangle^2}
 \Delta(\sin^2(\delta P))\,.
\end{eqnarray*}
The latter allows us to write $\langle\hat{T}\rangle=0$ as
\begin{equation}
 \langle\widehat{\sin(\delta P)}\rangle^2+ \langle\widehat{\cos(\delta
   P)}\rangle^2= \frac{1}{1+\langle\widehat{\cos(\delta P)}\rangle^{-2}
   \Delta(\sin^2(\delta P))}\,.
\end{equation}

\subsection{Complex moments}

The moment $\Delta(p_{\lambda}Q)$ is initially defined as a kinematical
moment, in which $\hat{p}_{\lambda}$ and $\hat{Q}$ need not be ordered
symmetrically in an explicit way because these are independent and commuting
operators. Once we transition to physical states by solving effective
constraints, we expect the ordering to become important because
$\widehat{Q\sin(\delta P)}$ does not commute with $\hat{Q}$. We start
with
\[
 \Delta(Qp_{\lambda})= \langle\hat{Q}\hat{p}_{\lambda}\rangle-
 \langle\hat{Q}\rangle \langle\hat{p}_{\lambda}\rangle
\]
as a kinematical moment, which becomes
\begin{equation} \label{Qpcomplex}
 \Delta(Qp_{\lambda})= \frac{1}{2}
 \langle\hat{Q}\hat{p}_{\lambda}+\hat{p}_{\lambda}\hat{Q}\rangle-
 \langle\hat{Q}\rangle \langle\hat{p}_{\lambda}\rangle +\frac{1}{2}\langle
 [\hat{Q},\hat{p}_{\lambda}]\rangle 
\end{equation}
in an explicit symmetric ordering. For physical moments involving
$p_{\lambda}$, $\hat{p}_{\lambda}$ must be ordered to the right so that we can
substitute $\hat{p}_{\lambda}\psi=\delta^{-1} \widehat{Q\sin(\delta P)}\psi$
on physical states. (If $\hat{p}_{\lambda}$ appears on the left, one would
have to use adjointness relations which depend on the Hilbert space used,
kinematical or physical, and cannot be taken for granted at the effective
level because $\widehat{Q\sin(\delta P)}$ is not a basic operator.)  The
commutator in (\ref{Qpcomplex}) then provides the imaginary contribution found
in (\ref{ComplexMoments}). 

While $\Delta(Qp_{\lambda})$ as the covariance of two independent variables is
no longer defined for physical states, one can interpret it as a composite
moment in which $\hat{p}_{\lambda}$ is understood as the operator $\delta^{-1}
\widehat{Q\sin(\delta P)}$. There should then be a physical moment for it,
which, unlike (\ref{ComplexMoments}) must be real. A simple guess suggests
that the physical moment is just the real part of (\ref{ComplexMoments}), or
\begin{equation}
 \Delta_{\rm phys}(Qp_{\lambda})= \langle\hat{Q}\rangle\Delta_{\rm
   phys}(Q\sin(\delta P))+ \langle\widehat{\sin(\delta P)}\rangle 
 \Delta_{\rm phys}(Q^2)\,.
\end{equation}
One can confirm this guess by computing $\Delta_{\rm phys}(Qp_{\lambda})$ for
states annihilated by the constraint, that is after substituting
$\frac{1}{2}(\hat{Q}\widehat{\sin(\delta P)}+\widehat{\sin(\delta P)}\hat{Q})$ for
$\delta\hat{p}_{\lambda}$ and symmetrizing $\hat{Q}$ and $\hat{p}_{\lambda}$:
\begin{eqnarray}
 \delta\Delta_{\rm phys}(Qp_{\lambda})&=&
 \frac{1}{2}\delta\langle\hat{Q}\hat{p}_{\lambda}+\hat{p}_{\lambda}\hat{Q}\rangle-
\delta \langle\hat{Q}\rangle \langle\hat{p}_{\lambda}\rangle\nonumber\\
&=& \frac{1}{4}\langle\hat{Q}\widehat{\sin(\delta P)}\hat{Q}+
\widehat{\sin(\delta P)}\hat{Q}^2+ \hat{Q}^2\widehat{\sin(\delta P)}+
\hat{Q}\widehat{\sin(\delta P)}\hat{Q}\rangle\nonumber\\
&&- \frac{1}{2}
 \langle\hat{Q}\rangle
 \langle\hat{Q}\widehat{\sin(\delta P)}+\widehat{\sin(\delta P)}\hat{Q}\rangle\,.
\end{eqnarray}
The ordering obtained in this way is not Weyl, but one can rearrange so that
\[
 \frac{1}{4}\left(\hat{Q}\widehat{\sin(\delta P)}\hat{Q}+
\widehat{\sin(\delta P)}\hat{Q}^2+ \hat{Q}^2\widehat{\sin(\delta P)}+
\hat{Q}\widehat{\sin(\delta P)}\hat{Q}\right)=
(\hat{Q}^2\widehat{\sin(\delta P)}^2)_{\rm Weyl}- \frac{1}{12} \hbar^2\delta^2
\widehat{\sin(\delta P)}\,.
\]
Moreover, for any third-order moment of the form $\Delta(A^2B)$ we have
\[
 \Delta(Q^2\sin(\delta P))= \langle(\hat{Q}^2\widehat{\sin(\delta P)}^2)_{\rm
   Weyl}\rangle- 2\langle\hat{Q}\rangle \Delta(Q\sin(\delta P))- \langle
 \widehat{\sin(\delta P)}\rangle \Delta(Q^2)-
 \langle\hat{Q}\rangle^2\langle\widehat{\sin(\delta P)}\rangle\,.
\]
Applying this for one term in our physical moment, we obtain
\begin{equation}
  \delta\Delta_{\rm phys}(Qp_{\lambda})= \Delta_{\rm phys}(Q^2\sin(\delta P))+
  \langle\hat{Q}\rangle\Delta_{\rm phys}(Q\sin(\delta P))+
  \langle\widehat{\sin(\delta P)}\rangle 
  \Delta_{\rm phys}(Q^2)- \frac{1}{2} \hbar^2\delta^2
  \langle\widehat{\sin(\delta P)}\rangle\,, 
\end{equation}
which indeed agrees with the real part of (\ref{ComplexMoments}) within the
present approximation, in which the third-order moment $\Delta_{\rm
  phys}(Q^2\sin(\delta P))=O(\hbar^{3/2})$ and the re-ordering term of order
$\hbar^2$ are small compared with the second-order moments of order $\hbar$. A
similar result holds for $\Delta_{\rm phys}(p_{\lambda}^2)$, which can be shown to be
\begin{eqnarray}
 \delta^2\Delta_{\rm phys}(p_{\lambda}^2) &=& \Delta_{\rm phys}(Q^2\sin^2(\delta P))-
 \Delta_{\rm phys}(Q\sin(\delta P))^2\\
&&+
 \frac{1}{4}\hbar^2\delta^2\left(1-\frac{1}{3}(\Delta_{\rm phys}(\sin^2(\delta
   P))+\langle\widehat{\sin(\delta P)}\rangle^2)\right)\nonumber\\
&&+
 2\langle\hat{Q}\rangle \Delta_{\rm phys}(Q\sin^2(\delta P))+
 2\langle\widehat{\sin(\delta P)}\rangle \Delta_{\rm phys}(Q^2\sin(\delta
 P))\nonumber\\ 
&&+
 \langle\hat{Q}\rangle^2\Delta_{\rm phys}(\sin^2(\delta P))+ \langle\sin(\delta
 \hat{P})\rangle^2 \Delta_{\rm phys}(Q^2)+ 2\langle\hat{Q}\rangle
 \langle\widehat{\sin(\delta P)}\rangle \Delta_{\rm phys}(Q\sin(\delta P))
\nonumber
\end{eqnarray}
on physical states. Compared with these calculations that require explicit
re-orderings, deriving moments from effective constraints is much easier.

\subsection{Complex time}

So far, in this appendix, we have not used deparameterization to distinguish
one of the phase-space variables (or an expectation value) as time. Doing so
sheds more light on the complex nature of moments, as well as the time
expectation value.

\subsubsection{Scalar time}
\label{a:ScalarTime}

If we choose $\lambda$ as time, as in the main body of this article,
expectation values and moments involving $\lambda$ and $p_{\lambda}$ are no
longer independent of those of $Q$ and $P$. Relationships for some of them are
provided by effective constraints, as discussed. Other moments, especially
those involving $\lambda$ are not even defined for physical states when
deparameterization is used, just as there is no strict time fluctuation in
quantum mechanics. Moments involving only $Q$ and $P$, on the other hand, are
physical and related to Dirac observables of the effective constrained system,
as shown in \cite{EffCons,EffConsRel}. Moments involving $p_{\lambda}$ can be
obtained after solving the constraint for $\hat{p}_{\lambda}$ and substituting
in the moments, as done in the preceding subsection and in Section
\ref{s:FluctEn} (see Eq.~(\ref{Deltalambda})). 

Moments involving $\lambda$ are not physical because they are not invariant
under the gauge transformations generated by the effective constraints
\begin{equation}
 C_{\lambda} = 2\delta^2\langle\hat{p}_{\lambda}\rangle \Delta(\lambda
 p_{\lambda})+ i\hbar\delta^2 \langle\hat{p}_{\lambda}\rangle-
 2\langle\hat{Q}\rangle \langle\widehat{\sin(\delta P)}\rangle^2 \Delta(\lambda Q)-
 2\langle\hat{Q}\rangle^2 \langle\widehat{\sin(\delta P)}\rangle
 \Delta(\lambda\sin(\delta P))
\end{equation}
and (\ref{Cpl}), (\ref{CQ}), (\ref{CP}). Following
\cite{EffTime,EffTimeLong}, one can see that the conditions
\begin{equation}
 0=\phi_1:=\Delta(\lambda^2)\quad,\quad 0=\phi_2:=\Delta(\lambda Q)\quad,\quad
 0=\phi_3:=\Delta(\lambda\sin(\delta P))
\end{equation}
are good gauge-fixing conditions for the first-order constraints
$C_{\lambda}$, $C_{p_{\lambda}}$, $C_Q$ and $C_{\sin(\delta P)}$. (Since
second-order moments have a Poisson structure that is not invertible, three
gauge-fixing conditions suffice for four first-class constraints. See
\cite{brackets} for constraints on non-symplectic spaces.) We can then
immediately solve the constraint $C_{\lambda}=0$ to obtain $\Delta(\lambda
p_{\lambda})=-\frac{1}{2}i\hbar$. (As a non-physical moment evaluated for
physical states, this moment need not be real. The complex value ensures that
the uncertainty relation is valid even with $\Delta\lambda=0$.)

After imposing the gauge-fixing conditions $\phi_i=0$, only one constraint is
left to the orders considered, along with a corresponding gauge flow that
amounts to evolution in internal time $\lambda$. The gauge generator
consistent with the gauge fixing is a linear combination $N^{\alpha}C_{\alpha}$
of the four constraints which have not yet been solved (all but $C_{\lambda}$)
so that $\{\phi_i,C_{\alpha}\}N^{\alpha}=0$ for $i=1,2,3$. Computing the
matrix $\{\phi_i,C_{\alpha}\}$ and solving for the restricted components of
$N^{\alpha}$, we obtain the combination
\begin{eqnarray*}
 N^{\alpha}C_{\alpha}&=& N\left(C-\frac{1}{2\langle\hat{p}_{\lambda}\rangle}
   C_{p_{\lambda}}- \frac{1}{2\langle\hat{Q}\rangle} C_Q-
   \frac{1}{2\langle\widehat{\sin(\delta P)}\rangle} C_{\sin(\delta P)}\right)\\
 &=& N\left(\delta^2\langle\hat{p}_{\lambda}\rangle^2- \langle\hat{Q}\rangle^2
   \langle\widehat{\sin(\delta P)}\rangle^2 -2\langle\hat{Q}\rangle
   \langle\widehat{\sin(\delta P)}\rangle \Delta(Q\sin(\delta P))\right)
\end{eqnarray*}
with only one free multiplier $N$. A convenient choice is
$N=1/(\delta\langle\hat{p}_{\lambda}\rangle)$, which gives us the
$\lambda$-Hamiltonian
\begin{equation} 
 H_{\lambda} \approx \langle\hat{p}_{\lambda}\rangle- \langle\hat{Q}\rangle
 \langle\widehat{\sin(\delta P)}\rangle- \Delta(Q\sin(\delta P))\,.
\end{equation}
(We have used the constraint $C=0$ and ignored products of second-order
moments.) 

One can easily solve some of the equations of motion to leading order,
ignoring moments. For instance, we have
\begin{equation}
 \frac{{\rm d}\langle\widehat{\sin(\delta P)}\rangle}{{\rm d}\lambda} =
 \langle\widehat{\sin(\delta P)}\rangle \langle\widehat{\cos(\delta P)}\rangle
 \quad,\quad \frac{{\rm d}\langle\widehat{\cos(\delta P)}\rangle}{{\rm
     d}\lambda}= - \langle\widehat{\sin(\delta P)}\rangle^2\,,
\end{equation}
so that $\langle\widehat{\sin(\delta P)}\rangle^2+ \langle\widehat{\cos(\delta
  P)}\rangle^2=: c$ is conserved. (This constant equals
$c=1-\Delta(\sin^2(\delta P))-\Delta(\cos^2(\delta P))\sim 1$ by the
Casimir constraint.) Using $c$,
we can decouple the two equations, and solve for
\begin{equation}
 \langle\widehat{\cos(\delta P)}\rangle(\lambda) = \sqrt{c}
 \tanh(-2\sqrt{c}(\lambda-\lambda_0))\,.
\end{equation}
With this solution, we obtain
\begin{equation}
 \langle\widehat{\sin(\delta P)}\rangle(\lambda)=
 \frac{\sqrt{c}}{\cosh(-2\sqrt{c}(\lambda-\lambda_0))}\,.
\end{equation}
We can then use the constraint $H_{\lambda}=0$ to find 
\begin{equation}
\langle\hat{Q}\rangle(\lambda) =
\frac{\langle\hat{p}_{\lambda}\rangle}{\sqrt{c}}
\cosh(-2\sqrt{c}(\lambda-\lambda_0))  \,.
\end{equation}
The expectation value of $\hat{Q}$ (the volume for $x=-1/2$) is bounded from
below by $\langle\hat{Q}\rangle_{\rm min}=
\langle\hat{p}_{\lambda}\rangle+O(\Delta(\cdot))$. (The solution for
$\langle\hat{Q}\rangle$ is exact for the harmonic ordering \cite{BouncePert}.)

\subsubsection{Curvature time}

Although the system is not deparameterizable by $P$, effective constraints can
be evaluated with local internal times valid only for a finite range. To this
end, we follow the same procedure as before but choose gauge-fixing conditions
\begin{equation} \label{GaugeP}
 0=\phi_1:=\Delta(\sin^2(\delta P))\quad,\quad 0=\phi_2:=\Delta(\lambda
 \sin(\delta P))\quad,\quad
 0=\phi_3:=\Delta(p_{\lambda}\sin(\delta P))\,.
\end{equation}
The $P$-Hamiltonian which preserves the new gauge-fixing conditions is now
\begin{equation}
 H_{\sin(\delta P)} = N\left(\delta^2\langle\hat{p}_{\lambda}\rangle^2-
   \langle\hat{Q}\rangle^2 \langle\widehat{\sin(\delta P)}\rangle^2- i\hbar
   \delta\langle\hat{Q}\rangle \langle\widehat{\sin(\delta P)}\rangle
   \langle\widehat{\cos(\delta P)}\rangle \right)\,,
\end{equation}
whose imaginary contribution comes from the complex $\Delta(Q\sin(\delta
P))=\frac{1}{2} i\hbar\delta\langle\widehat{\cos(\delta P)}\rangle$ (a
non-physical moment when $P$ is chosen as time). The $P$-Hamiltonian can be
real only if the time expectation value $\langle\widehat{\sin(\delta
  P)}\rangle$ is complex. (The expectation value
$\langle\hat{p}_{\lambda}\rangle$ is physical in this choice of time and must
be real.) Since the imaginary contribution to $H_{\sin(\delta P)}$ is of the
order of $\hbar\delta$ and therefore small, the imaginary contribution to
$\langle\widehat{\sin(\delta P)}\rangle$ is small and we can expand
\begin{equation}
 \langle\widehat{\sin(\delta P)}\rangle = {\rm Re}\langle\widehat{\sin(\delta
   P)} \rangle+i\delta 
 \langle\widehat{\cos(\delta P)}\rangle I
\end{equation}
as well as
\begin{equation}
 \langle\widehat{\cos(\delta P)}\rangle = {\rm Re}\langle\widehat{\cos(\delta 
 P)}\rangle-i\delta 
 \langle\widehat{\sin(\delta P)}\rangle I
\end{equation}
with $I$ playing the role of an imaginary part of $\langle\hat{P}\rangle$.
The $P$-Hamiltonian is then real provided that 
\begin{eqnarray*}
&& \langle\widehat{\sin(\delta P)}\rangle^2+ i\hbar
   \delta \frac{\langle\widehat{\sin(\delta P)}\rangle
   \langle\widehat{\cos(\delta P)}\rangle}{\langle\hat{Q}\rangle} \sim
 ({\rm Re}\langle\widehat{\sin(\delta P)}\rangle)^2\\
&&+ 2i\delta {\rm Re}\langle\widehat{\sin(\delta P)}\rangle
   {\rm Re}\langle\widehat{\cos(\delta P)}\rangle I+
   i\hbar\delta \frac{{\rm Re}\langle\widehat{\sin(\delta P)}\rangle
  {\rm Re} \langle\widehat{\cos(\delta P)}\rangle}{\langle\hat{Q}\rangle}
+O(I^2) 
\end{eqnarray*}
has a vanishing imaginary part. Thus, $I=
-\frac{1}{2} \hbar/\langle\hat{Q}\rangle$.

We finally use this example to show that relationships between moments depend
on the choice of internal time, or on the deparameterization scheme and the
physical Hilbert space one may construct. With the curvature parameter $P$
chosen as time and the gauge fixing (\ref{GaugeP}), the general relation
\[
\delta^2\Delta(p_{\lambda}^2)=\langle\hat{Q}\rangle^2\Delta(\sin^2(\delta P))+
\langle\widehat{\sin(\delta P)}\rangle^2 \Delta(Q^2)+ 2\langle\hat{Q}\rangle
\langle\widehat{\sin(\delta P)}\rangle \Delta(Q\sin(\delta P))
\]
reduces to
\[
 \delta^2\Delta(p_{\lambda}^2)=
 \langle\widehat{\sin(\delta P)}\rangle^2 \Delta(Q^2)+ i\hbar\delta
 \langle\hat{Q}\rangle \langle\widehat{\sin(\delta P)}\rangle
 \langle\widehat{\cos(\delta P)}\rangle\,.
\]
The complex contribution is no longer surprising. The real part of this
equation implies
\[
\frac{\Delta_{\rm phys} p_{\lambda}}{\langle\hat{p}_{\lambda}\rangle} =
\frac{\Delta_{\rm phys} Q}{\langle\hat{Q}\rangle}
\]
for all states with the present choice of internal time,
in contrast to the state-dependent and more-complicated behavior seen in the
main body of this paper for $\lambda$ or $\phi$ as internal time. This result
serves as a reminder that the time variable and its momentum are not defined
for physical states, neither as expectation values nor in moments. Moments of
the momentum can be defined only indirectly upon using the constraint.

\end{appendix}


\end{document}